\documentclass[11pt]{article}
\usepackage[margin=1in]{geometry}
\usepackage{amsmath,amssymb,amsthm,booktabs,array}
\usepackage{graphicx}
\usepackage[round]{natbib}
\usepackage{setspace}
\usepackage{placeins}
\usepackage{xurl}
\usepackage[hidelinks]{hyperref}
\hypersetup{
  pdftitle={Prizes, Shared Hazard, and the Speed of Frontier AI},
  pdfauthor={Li Gan}
}
\newtheorem{proposition}{Proposition}

\newcommand{\dvN}{3}
\newcommand{\dvPiShare}{50}
\newcommand{\dvPrizeOverB}{3.0}
\newcommand{\dvKappa}{0.075}
\newcommand{\dvKappaYearOdds}{2.5}

\newcommand{\dvVRacePct}{8.3}
\newcommand{\dvVRaceOverR}{33}
\newcommand{\dvRMonthPct}{0.25}
\newcommand{\dvRiskRace}{75}
\newcommand{\dvXCoop}{0.06}
\newcommand{\dvTCoop}{69}

\newcommand{\dvRiskCoop}{6}
\newcommand{\dvXNoContest}{0.33}
\newcommand{\dvRiskNoContest}{12}
\newcommand{\dvTNoContest}{55}

\newcommand{\dvLabsTwoX}{11}
\newcommand{\dvLabsTwoT}{27.7}
\newcommand{\dvLabsTwoRisk}{53}

\newcommand{\dvLabsFourX}{95}
\newcommand{\dvLabsFourT}{14.6}
\newcommand{\dvLabsFourRisk}{81}

\newcommand{\dvLeadCoopX}{7.7}
\newcommand{\dvLeadCoopT}{30}
\newcommand{\dvLeadCoopRisk}{47}

\newcommand{\dvKappaHalfX}{8}
\newcommand{\dvKappaHalfT}{29.4}
\newcommand{\dvKappaHalfRisk}{49}
\newcommand{\dvKappaDoubleX}{95}
\newcommand{\dvKappaDoubleT}{14.6}
\newcommand{\dvKappaDoubleRisk}{81}

\newcommand{\dvUniCoopOthers}{35}
\newcommand{\dvUniCoopRisk}{57}
\newcommand{\dvUniCoopCommon}{6}
\newcommand{\dvUniCoopShare}{27}

\newcommand{\dvUniFiveOthers}{44}
\newcommand{\dvUniFiveRisk}{65}
\newcommand{\dvUniFiveCommon}{40}
\newcommand{\dvUniFiveShare}{28}

\newcommand{\dvUniTenOthers}{46}
\newcommand{\dvUniTenRisk}{68}
\newcommand{\dvUniTenCommon}{52}
\newcommand{\dvUniTenShare}{30}

\newcommand{\rrTauZero}{11.2}
\newcommand{\rrTauPoint}{0.11}
\newcommand{\rrHalfT}{22.1}
\newcommand{\rrNoContestHurdle}{0.75}
\newcommand{\rrExternality}{0.50}
\newcommand{\rrContestPart}{7.5}

\newcommand{\rrHurdleTwo}{5.0}
\newcommand{\rrHurdleFour}{10.6}
\newcommand{\rrMarginalShare}{94}
\newcommand{\rrYearHeadWin}{55}
\newcommand{\rrProbShare}{91}
\newcommand{\rrGapContest}{63}
\newcommand{\rrGapCoord}{69}
\newcommand{\rrSensTargetLow}{65.9}
\newcommand{\rrSensTargetHigh}{79.3}
\newcommand{\rrSensPrizeRisk}{48.7}
\newcommand{\rrSensPrizeShare}{86.1}
\newcommand{\rrSensPeak}{0.679}

\newcommand{\lssRacePFive}{75.0}

\newcommand{\lssRaceXTwentyFive}{75.0}
\newcommand{\lssRacePTwentyFive}{53.6}
\newcommand{\lssCoopPTwentyFive}{4.1}
\newcommand{\lssShareTwentyFive}{92}
\newcommand{\dcHurdleNoExt}{2.75}
\newcommand{\dcXNoExt}{3.1}
\newcommand{\dcRiskNoExt}{33}
\newcommand{\dcContestSecond}{39}

\newcommand{\dcContestShapley}{65}

\newcommand{\sgRacePThreeQ}{89}

\newcommand{\sgCoopXBase}{0.06}
\newcommand{\sgCoopPBase}{5.8}

\newcommand{\sgFullPOneQ}{63}
\newcommand{\sgRaceXOneQ}{61}
\newcommand{\sgRacePOneQ}{66}

\newcommand{\sgCoopXOneQ}{0.22}
\newcommand{\sgCoopPOneQ}{4.9}

\newcommand{\sgGivenBackMin}{26}
\newcommand{\sgGivenBackMax}{34}
\newcommand{\sgShareMin}{91}
\newcommand{\sgShareMax}{92}
\newcommand{\sgPeakThreeQ}{74}
\newcommand{\sgVDropCoop}{33}
\newcommand{\sgVDropRace}{2}
\newcommand{\tkTakeoffEighteen}{18}
\newcommand{\tkTakeoffThirtySix}{36}
\newcommand{\tkComputeFoldEighteen}{8}
\newcommand{\tkComputeFoldThirtySix}{64}
\newcommand{\tkSigmaEighteen}{0.086}
\newcommand{\tkSigmaThirtySix}{0.084}
\newcommand{\tkKappaEighteen}{0.075}
\newcommand{\tkKappaThirtySix}{0.096}
\newcommand{\tkRiskRaceEighteen}{75}
\newcommand{\tkRiskRaceThirtySix}{75}
\newcommand{\tkVRacePctEighteen}{8.3}
\newcommand{\tkVRacePctThirtySix}{10.4}
\newcommand{\tkXCoopEighteen}{0.06}
\newcommand{\tkXCoopThirtySix}{0.15}
\newcommand{\tkTCoopEighteen}{69}
\newcommand{\tkTCoopThirtySix}{84}
\newcommand{\tkRiskCoopEighteen}{6}
\newcommand{\tkRiskCoopThirtySix}{4}
\newcommand{\tkXNoContestEighteen}{0.33}
\newcommand{\tkXNoContestThirtySix}{0.76}
\newcommand{\tkRiskNoContestEighteen}{12}
\newcommand{\tkRiskNoContestThirtySix}{10}
\newcommand{\tkGapContestEighteen}{63}
\newcommand{\tkGapContestThirtySix}{65}
\newcommand{\tkContestShapleyEighteen}{65}
\newcommand{\tkContestShapleyThirtySix}{69}
\newcommand{\tkLeadCoopXEighteen}{7.7}
\newcommand{\tkLeadCoopXThirtySix}{10.9}
\newcommand{\tkLeadCoopTEighteen}{30}
\newcommand{\tkLeadCoopTThirtySix}{48}
\newcommand{\tkLeadCoopRiskEighteen}{47}
\newcommand{\tkLeadCoopRiskThirtySix}{43}

\newcommand{\tkGapCoordThirtySix}{71}

\newcommand{\lswKappa}{0.075}
\newcommand{\lswSwitchTzero}{13}
\newcommand{\lswSwitchPreg}{49}
\newcommand{\lswSwitchOmega}{0.79}
\newcommand{\lswExpoNine}{6.4}
\newcommand{\lswExpoEighteen}{2.6}
\newcommand{\lswKnowNine}{11.0}
\newcommand{\lswKnowEighteen}{10.8}
\newcommand{\lswPeakNine}{16}
\newcommand{\lswHeldFirstMin}{91}
\newcommand{\lswHeldFirstMax}{92}
\newcommand{\lswHeldShapMin}{65}
\newcommand{\lswHeldShapMax}{66}

\newcommand{\lswXRaceThirtySix}{33}
\newcommand{\lswSweepMin}{69}
\newcommand{\lswSweepMax}{96}
\newcommand{\lswPfullHalf}{86}
\newcommand{\lswPfullQuarter}{69}
\newcommand{\lswNFive}{83}
\newcommand{\lswDoublingTwenty}{89}
\newcommand{\lswPregHalf}{89}
\newcommand{\lswShapPfullQuarter}{44}
\newcommand{\lswSigmaNine}{0.091}
\newcommand{\lswXCoopNine}{0.02}
\newcommand{\lswTCoopNine}{63}
\newcommand{\lswRiskCoopNine}{9.8}
\newcommand{\lswRiskNoContestNine}{15.6}
\newcommand{\lswHeldXNine}{95}
\newcommand{\lswHeldTNine}{7.0}
\newcommand{\lswHeldRiskNine}{77}
\newcommand{\lswHeldFirstNine}{91}
\newcommand{\lswHeldShapNine}{66}
\newcommand{\lswFitKappaNine}{--}
\newcommand{\lswFitRiskNine}{--}
\newcommand{\lswFitFirstNine}{--}
\newcommand{\lswFitShapNine}{--}
\newcommand{\lswSigmaTwelve}{0.089}
\newcommand{\lswXCoopTwelve}{0.03}
\newcommand{\lswTCoopTwelve}{65}
\newcommand{\lswRiskCoopTwelve}{7.9}
\newcommand{\lswRiskNoContestTwelve}{13.8}
\newcommand{\lswHeldXTwelve}{95}
\newcommand{\lswHeldTTwelve}{9.4}
\newcommand{\lswHeldRiskTwelve}{79}
\newcommand{\lswHeldFirstTwelve}{92}
\newcommand{\lswHeldShapTwelve}{65}
\newcommand{\lswFitKappaTwelve}{--}
\newcommand{\lswFitRiskTwelve}{--}
\newcommand{\lswFitFirstTwelve}{--}
\newcommand{\lswFitShapTwelve}{--}
\newcommand{\lswSigmaEighteen}{0.086}
\newcommand{\lswXCoopEighteen}{0.06}
\newcommand{\lswTCoopEighteen}{69}
\newcommand{\lswRiskCoopEighteen}{5.8}
\newcommand{\lswRiskNoContestEighteen}{11.8}
\newcommand{\lswHeldXEighteen}{50}
\newcommand{\lswHeldTEighteen}{18.0}
\newcommand{\lswHeldRiskEighteen}{75}
\newcommand{\lswHeldFirstEighteen}{91}
\newcommand{\lswHeldShapEighteen}{65}
\newcommand{\lswFitKappaEighteen}{0.075}
\newcommand{\lswFitRiskEighteen}{75}
\newcommand{\lswFitFirstEighteen}{91}
\newcommand{\lswFitShapEighteen}{65}
\newcommand{\lswSigmaTwentyFour}{0.085}
\newcommand{\lswXCoopTwentyFour}{0.09}
\newcommand{\lswTCoopTwentyFour}{74}
\newcommand{\lswRiskCoopTwentyFour}{4.8}
\newcommand{\lswRiskNoContestTwentyFour}{10.7}
\newcommand{\lswHeldXTwentyFour}{37}
\newcommand{\lswHeldTTwentyFour}{26.0}
\newcommand{\lswHeldRiskTwentyFour}{70}
\newcommand{\lswHeldFirstTwentyFour}{91}
\newcommand{\lswHeldShapTwentyFour}{66}
\newcommand{\lswFitKappaTwentyFour}{0.085}
\newcommand{\lswFitRiskTwentyFour}{75}
\newcommand{\lswFitFirstTwentyFour}{92}
\newcommand{\lswFitShapTwentyFour}{67}
\newcommand{\lswSigmaThirtySix}{0.084}
\newcommand{\lswXCoopThirtySix}{0.15}
\newcommand{\lswTCoopThirtySix}{84}
\newcommand{\lswRiskCoopThirtySix}{3.8}
\newcommand{\lswRiskNoContestThirtySix}{9.6}
\newcommand{\lswHeldXThirtySix}{33}
\newcommand{\lswHeldTThirtySix}{39.2}
\newcommand{\lswHeldRiskThirtySix}{66}
\newcommand{\lswHeldFirstThirtySix}{91}
\newcommand{\lswHeldShapThirtySix}{66}
\newcommand{\lswFitKappaThirtySix}{0.096}
\newcommand{\lswFitRiskThirtySix}{75}
\newcommand{\lswFitFirstThirtySix}{92}
\newcommand{\lswFitShapThirtySix}{69}
\newcommand{\hzFloor}{11.8}
\newcommand{\hzVTen}{0.63}
\newcommand{\hzVFive}{0.31}
\newcommand{\hzFirstTwenty}{58}
\newcommand{\hzFirstFifteen}{35}
\newcommand{\hzrVHi}{8.3}
\newcommand{\hzrMultHi}{33}
\newcommand{\hzrRaceHi}{75}
\newcommand{\hzrCoopHi}{5.8}
\newcommand{\hzrXCoopHi}{0.06}
\newcommand{\hzrXNoContestHi}{0.33}
\newcommand{\hzrXNoExtHi}{3.1}
\newcommand{\hzrFirstHi}{91}
\newcommand{\hzrSecondHi}{39}
\newcommand{\hzrShapHi}{65}
\newcommand{\hzrGapHi}{69}
\newcommand{\hzrVLo}{1.7}
\newcommand{\hzrMultLo}{6.9}
\newcommand{\hzrRaceLo}{25}
\newcommand{\hzrCoopLo}{3.8}
\newcommand{\hzrXCoopLo}{0.72}
\newcommand{\hzrXNoContestLo}{5.2}
\newcommand{\hzrXNoExtLo}{3.1}
\newcommand{\hzrFirstLo}{69}
\newcommand{\hzrSecondLo}{20}
\newcommand{\hzrShapLo}{44}
\newcommand{\hzrGapLo}{21}
\newcommand{\hzrRegHi}{40}
\newcommand{\hzrRegLo}{10}

\title{Prizes, Shared Hazard, and the Speed of Frontier AI}
\author{Li Gan\thanks{Department of Economics, Texas A\&M University. Email: \url{ganli@tamu.edu}. I thank Eric Gan and seminar participants for helpful discussions. AI software (Anthropic's Claude) was used under the author's direction to derive and check the algebra, write the calibration and simulation code, and edit the draft. All output was checked by the author, who is responsible for the content. An interactive companion page, which runs the paper's model in the browser, is at \url{https://ligan98.github.io/frontier-pacing/en/labs.html}.}}
\date{November 2026}

\begin{document}
\maketitle

\begin{abstract}
Frontier AI labs compete in a contest for commercial leadership while sharing the risk of catastrophe. Each lab gains from arriving earlier but bears only part of the hazard its speed adds, which this paper calls the hazard externality. In an illustrative calibrated takeoff model, three racing labs get through the takeoff period in 18 months, while a common rule on research shares would stretch it to \dvTCoop{} months. Neither force alone explains racing that fast. Removing the contest first closes over ninety percent of the gap in catastrophe probability between racing and a common rule when the catastrophe probability at full speed is 75 percent, and \hzrFirstLo{} percent when it is 25 percent; averaged over both orders of removal, the contest explains \dcContestShapley{} and \hzrShapLo{} percent. More labs make each lab race harder, when industry hazard at a common research share does not depend on the number of labs. Under open source in its ideal form, where every lab publishes its capability findings at once, labs slow to the coordinated pace, because no lab can get ahead and each bears all the hazard its research adds. One lab's restraint achieves much less than a common rule. The White House Accord on Super Intelligence of September 2026 audits safety but sets no limit on speed. In the model it does not slow the race, because each lab still gains as much from speeding up.
\end{abstract}

\vspace{1em}
\noindent Keywords: Artificial intelligence, technology racing, patent races, existential risk, common rules.\\
\noindent JEL Classification: D43, L13, O32, O33.

\newpage

\section{Introduction}

How much faster do competing frontier AI labs develop than they would under a common rule, and why? In the paper's calibration, three racing labs get through the takeoff period in 18 months, while a common rule would stretch it to \dvTCoop{} months. Discounting and the hazard externality alone cannot explain why racing labs go that fast, so the contest for the prize supplies the rest. Proposals to restrain frontier development face an objection from inside the industry: a lab that slows capability research may lose commercial leadership to a rival \citep{amodei2026,aifutures2026}.

The latest industry response is voluntary. On September 29, 2026, six leading AI and chip companies, Google, Anthropic, Meta, OpenAI, xAI and Nvidia, signed the White House Accord on Super Intelligence. It commits them to four layers of safety controls and audits \citep{whitehouse2026}. It sets no penalties and does not address how fast labs develop. In the model, the accord will not slow the race, because it leaves each lab's research share free. Its audits can lower risk through safety effort, but each lab still gains as much from speeding up.\footnote{Table~\ref{tab:safety} shows the limit. Even a 25 percent safety share imposed on every lab lowers racing risk from \lssRacePFive{} to \lssRacePTwentyFive{} percent, against \lssCoopPTwentyFive{} percent under coordination.} Section~\ref{sec:accord} applies the model to the accord.

Each lab chooses a stationary share of compute for capability research; safety effort is fixed and customers receive the rest. Customer services do not enter the labs' terminal payoff, so the game isolates the incentive to develop faster. A catastrophe originating at any frontier destroys every lab's terminal stake, while a logit contest over arrival dates allocates a commercial prize. The benchmark is common-share coordination, which chooses one research share for every lab to maximize their joint terminal payoff. We also call it a common rule when it binds every lab. It differs from a social optimum because consumer surplus and losses outside the labs' stakes do not enter that objective.

The paper has four results.

First, the contest and the hazard externality raise the racing hurdle above the industry's hurdle under coordination (Proposition~\ref{prop:race}). A hurdle is the hazard reduction that makes a lab accept one more month of delay. Each lab counts only its share of the hazard its speed adds, while keeping the full discounting benefit and better odds of winning the prize. Coordination counts the full industry hazard, and common arrival leaves the prize unchanged. The gap between the two hurdles is the racing wedge. We call the hazard saved by an additional month of delay the value of a month. When that value rises with research, the higher racing hurdle means faster development.

The contest supplies most of the racing wedge in the baseline. Contest sensitivity is fitted so that racing labs put 50 percent of compute into research. At that share the value of a month is \dvVRacePct{} percent a month, but discounting and the hazard externality together supply only \rrNoContestHurdle{} percent a month. The contest must supply the rest.

Removing the contest first closes over ninety percent of the baseline catastrophe-probability gap between racing and coordination. This is a sequential counterfactual, so its size depends on the order in which the two forces are removed. Removing the hazard externality first gives the contest a smaller share, because counting only one $N$th of the hazard multiplies every private gain from speed by $N$, the contest's included. At a full-speed probability of 25 percent, the lower end of the range we consider, removing the contest first closes \hzrFirstLo{} percent of the gap and the average over the two orders is \hzrShapLo{} percent (Section~\ref{sec:hazard_level}). Section~\ref{sec:numbers} reports both orders and their average, keeping this probability comparison distinct from the marginal racing wedge.

Second, more labs raise each lab's racing hurdle (Proposition~\ref{prop:labs}). With a fixed hazard per frontier, research shares can fall as labs are added, but aggregate risk still rises in the calibrated cases.

Third, open source slows labs to the coordinated pace (Proposition~\ref{prop:publish}). Under open source in its ideal form, every lab publishes its capability findings at once, so all labs work from one shared frontier. No lab can then get ahead, which removes the contest. Each lab's research also moves every lab's frontier, so each lab bears all the hazard its research adds, which removes the hazard externality. Real releases fall short of this ideal, because they come with a lag, labs keep some research private, and published findings also reach foreign rivals. Monitoring how much compute goes to safety is a different policy, because it makes no finding public.

Fourth, one lab's restraint achieves much less than a common rule (Table~\ref{tab:uni}). When one lab adopts the coordinated share, its rivals slow only part of the way. The industry then gets only part of the risk reduction the same rule achieves when it binds every lab. Rivals' responses depend on the prize structure and need not reinforce restraint. A voluntary pledge on speed achieves little unless it binds every lab (Section~\ref{sec:unilateral}).

The numerical shares depend on the prize, the assumed racing speed, the full-speed probability, and the order of the counterfactuals. Appendix~\ref{app:calibration} varies those inputs and the takeoff technology. When the takeoff period is shorter than about \lswSwitchTzero{} months, no contest sensitivity supports the imposed racing share; keeping sensitivity fixed instead yields a faster race. The same failure occurs when the full-speed probability is below about \hzFloor{} percent. These checks distinguish a failure to fit the assumed speed from the contest ceasing to reward speed. Section~\ref{sec:scope} states the model's limits.

The analysis builds on patent races \citep{loury1979,dasgupta1980,fudenberg1982,harris1987} and on models of AI competition. \citet{armstrong2016} show how the value of arriving first can weaken safety incentives. \citet{bdmdp2026} show that fragmentation shifts a fixed pool of industry resources toward speed. \citet{fudenbergkoh2026} study two firms that choose how fast to advance toward a rising safety threshold under a common disaster hazard. Low risk sustains perpetual racing, high risk rules it out, and hidden capabilities make pacing harder to sustain. \citet{jensen2023} study compute prices and safety subsidies. We fix stationary research shares, split the racing hurdle into discounting, the hazard externality, and a contest part, and compute each on a calibrated takeoff period. Competitive underinvestment in safety and the compute--safety trade-off are familiar mechanisms; the contribution is the decomposition and its size.

The growth--risk and regulation literatures examine the broader social trade-offs \citep{jones2024,acemoglu2024,bostrom2026,binder2021}. Companion papers study single-lab commitment bounds, customer-versus-research compute allocation, and verified restraint between countries \citep{gan2026insights,gan2026allocation,gan2026agreements}.

Section~\ref{sec:model} presents the takeoff model and baseline. Section~\ref{sec:developers} derives the racing results, and Section~\ref{sec:numbers} quantifies them. Section~\ref{sec:accord} applies the model to the White House Accord on Super Intelligence. Section~\ref{sec:conclusion} discusses competition policy and the model's limits.

\section{The Takeoff Model}\label{sec:model}

\subsection{The Pieces in Words}

The model follows one AI frontier through a takeoff period. It is the takeoff model of \citet{gan2026insights} and \citet{gan2026allocation}, with three pieces: compute, capability, and safety knowledge. Compute powers research, capability affects its productivity, and safety knowledge reduces the danger of development.

Compute is the main input of a frontier lab. It means the chips, data centers, and electricity that a lab runs. A lab uses compute in three ways. It serves customers, which is called inference. It does capability research, which trains stronger models. And it does safety research, which makes models less likely to cause harm. The three uses draw on the same chips, so compute spent on one use is not available for the others.

Capability research speeds up as models improve, because AI models help with AI research. The takeoff period starts at the Automated Coder, a model that automates machine learning engineering. It ends at top-expert-dominating AI, a model that surpasses human scientists in every domain. Capability is an index that runs from 0 at the first milestone to 1 at the second.

The catastrophe is a catastrophic failure or a loss of control over AI. We read this danger as misalignment. A more capable model is more likely to pursue goals its developers did not intend and better able to escape control, so the danger rises with capability. The hazard measures the rate per month at which it occurs, given that it has not happened yet. Hazards add up over time. A hazard of 1 percent a month for a year gives a cumulative hazard of 0.12, or a probability of catastrophe of about 11 percent. Danger has two sources. State hazard is the danger of having a model of a given capability in use, and it accrues per month. Step hazard is the danger of training a stronger model whose properties nobody yet knows, and it accrues per unit of capability gained.

Safety research builds a stock of knowledge, and knowledge lowers both hazards at every later capability level. Knowledge learned now keeps protecting later, stronger models. So slowing down has a benefit and a cost. Each capability level is reached with more knowledge, but more months are spent at each level. Which effect is larger decides whether delay saves risk.

The three pieces interact through time. At full speed a lab puts half of its compute into capability research and 5 percent into safety, and the takeoff period takes 18 months \citep{aifutures2026}. Pacing means moving slower than this, by cutting the research share or by pausing. Installed compute keeps growing whatever the lab does. So a lab that slows down or pauses reaches each later capability level with more compute. A pause that stops capability research still lets safety research continue. Later sections add rivals to this single frontier.

\subsection{Compute and Capability}

Installed compute grows exponentially. Time $t \ge 0$ is measured in months, and total installed computing capacity $K(t)$ grows at rate $g_K$:
\begin{equation}
K(t) = K_0\,e^{g_K t}.
\end{equation}
The calibration sets $g_K=\ln2/6$, so compute doubles every six months. In each period, compute is divided among three uses: customer inference services $i(t)$, capability research $x(t)$, and safety research $s(t)$, satisfying the resource constraint
\begin{equation}
i(t) + x(t) + s(t) = 1, \qquad i(t), x(t), s(t) \ge 0.
\end{equation}
The shares sum to one, so compute moved into research comes out of customer services or safety.

Frontier capability is a state $F(t) \in [0, 1]$. At $F=0$, systems automate machine learning engineering (the Automated Coder milestone). At $F=1$, they surpass human scientific capabilities across all domains (top-expert-dominating AI, or TED-AI). Capability advances according to
\begin{equation}
\dot F(t) = \phi\bigl(F(t)\bigr)\,\bigl(x(t) K(t)\bigr)^\varepsilon,
\label{eq:capability_ode}
\end{equation}
where $\varepsilon \in (0, 1]$ is the research elasticity and $\phi(F)>0$ is the research productivity of a model of capability $F$, because more capable models improve the research process. So the pace of capability is research productivity times research capacity, the compute put into research raised to the research elasticity. The results hold for any continuous positive $\phi$. The calibration sets $\phi(F)=\theta e^{\rho F}$, so research at TED-AI is ten times as fast as at the Automated Coder for the same compute. At a fixed share, research capacity $(xK)^\varepsilon$ grows at the rate $\gamma=\varepsilon g_K$. With $\varepsilon=1/2$, quadrupling research compute only doubles the pace at a given capability. Research capacity grows at half the rate of compute, doubling every year, with $\gamma=\ln2/12$.

Progress accelerates during the full-speed takeoff period. Along an unconstrained full-speed path with baseline research share $x_0 = 0.50$ and safety share $s_0 = 0.05$, the takeoff period from $F=0$ to $F=1$ takes $T_0 = 18$ months. It reaches the halfway point, $F=1/2$, after about 15 months, so the second half of the capability index takes about 3 months.

\subsection{Safety Knowledge and Hazard}

Safety knowledge accumulates through dedicated safety research:
\begin{equation}
\dot S(t) = \sigma\bigl(F(t)\bigr)\,y\bigl(s(t)\bigr),
\label{eq:safety_ode}
\end{equation}
where $\sigma(F)$ is the safety yield, the knowledge a month of safety research adds at the full-speed share $s_0$. The yield is a function of capability, because stronger models change how productive safety research is. The learning multiplier $y(s)$ is continuous and increasing, with $y(s_0)=1$. The calibration sets $y(s)=\psi+(1-\psi)(s/s_0)^\alpha$ with $\psi=\alpha=1/2$: half of baseline learning needs no dedicated compute, and the other half has square-root returns in the safety share. Quadrupling the safety share doubles the second half, so learning runs 1.5 times as fast. With constant $\sigma$, safety knowledge accumulates at a constant calendar-time rate for a fixed safety share. In the calibration $\sigma$ is constant at 0.086 per month, so a year at the full-speed share adds about 1.03 units of knowledge. A capability pause permits further safety learning; it does not freeze all research.

At each moment the frontier generates an instantaneous hazard of catastrophic failure or loss of control, $\lambda(t)$. It depends on the capability level, the pace of capability advance, and accumulated safety knowledge:
\begin{equation}
\lambda(t) = \Bigl[ h\bigl(F(t)\bigr) + k\bigl(F(t)\bigr)\,\dot F(t) \Bigr]\,e^{-S(t)},
\label{eq:hazard}
\end{equation}
where $h(F)$ represents state hazard (the hazard of maintaining systems at capability $F$) and $k(F)\dot F$ represents step hazard (the hazard incurred from destabilizing transitions during rapid capability jumps). Knowledge scales both terms by $e^{-S}$. The year of safety research above multiplies the hazard at every later level by $e^{-1.03}$, about 0.36. During a pause, step hazard stops, while state hazard continues at the frozen capability and falls as knowledge grows. Cumulative hazard over the takeoff period is
\begin{equation}
H = \int_0^T \lambda(t)\,dt,
\end{equation}
and the cumulative probability of catastrophe during the takeoff period is $P = 1 - e^{-H}$. We call $H$ the cumulative hazard and $P$ the catastrophe probability. We count hazard only during the takeoff period, as the AI Futures Project's 75 percent does. So the model assumes that the danger ends at completion, for example because a top-expert-dominating AI still under control can help align its successors \citep{aifutures2026}. In the baseline calibration, unpaced development carries a catastrophe probability of $P=\dvRiskRace\%$, with state and step hazard contributing equally. That is a cumulative hazard of $\ln4$, about 1.39. The calibration puts three quarters of this hazard in the second half of the capability index, the last 3 months of a full-speed takeoff period.

\subsection{Baseline Numbers}

Table~\ref{tab:calibration} collects the illustrative baseline used below. The numbers describe a scenario; they are not estimates of any lab's capabilities or of catastrophe risk. Four of the targets come from the worked example of the AI Futures Project \citeyearpar{aifutures2026}. At full speed the takeoff period takes 18 months and carries a 75 percent probability of catastrophe. A safety-case regime, in which third-party assessors estimate each company's risk and the government caps it, cuts that probability to about 40 percent. The example has three frontier companies. We treat 75 percent as the upper end of a range and also report 25 percent, the lower case of Section~\ref{sec:hazard_level}. At 25 percent the safety-case regime's probability falls from \hzrRegHi{} to \hzrRegLo{} percent, so that the regime cuts cumulative hazard in the same proportion.

Removing the contest first closes most of the racing--coordination gap in every reported case. Appendix~\ref{app:calibration} repeats the main results with takeoff periods from 9 to 36 months and varies the other parameters one at a time (Tables~\ref{tab:takeoff36} and~\ref{tab:takeoff_range}, Figure~\ref{fig:labs_sweep}).

\begin{table}[!htb]
\centering\small
\caption{Scenario inputs, fitted parameters, and model implications}\label{tab:calibration}
\begin{tabular}{@{}p{.23\textwidth}p{.27\textwidth}>{\raggedright\arraybackslash}p{.44\textwidth}@{}}
\toprule
Parameter & Value & Calibration status\\
\midrule
$T_0$; $P_{\rm full}$ & 18 months; 0.75 & Scenario targets from \citet{aifutures2026}\\
$L_{\rm cal}$; $P_{\rm reg}$ & 12 months; 0.40 & Safety-case target from \citet{aifutures2026}, imposed on a stopped-clock pause\\
$\omega$; $q$ & 0.50; 0.75 & Assumed state and late-hazard shares\\
$\varepsilon$; $\rho$ & 0.50; $\ln 10$ & Assumed capability technology, $\phi(F)=\theta e^{\rho F}$\\
$g_K$; $\gamma$ & $\ln2/6$; $\ln2/12$ & Assumed compute growth; implied growth of research capacity\\
$x_0$; $s_0$ & 0.50; 0.05 & Imposed research; assumed safety shares\\
$\psi$; $\alpha$ & 0.50; 0.50 & Assumed learning multiplier $y(s)$\\
$\sigma$ & 0.086116 & Fitted to the 40\% target ($P_{\rm reg}$)\\
$b_0$; $\theta$ & 0.012348; 0.017463 & Fitted to the 18-month takeoff period ($T_0$); implied for $K_0=1$\\
$a_h$; $c_h$ & 0.380750; 2.197225 & Scale fitted to the 75\% target ($P_{\rm full}$); shape set by the assumed $q$\\
$N$; $\pi$ & 3; 0.50 & Companies as in \citet{aifutures2026}; assumed prize share\\
$\kappa$; $r$ & 0.075034; 0.0025 & Fitted so racing labs choose the imposed 50\% research share ($x_0$); assumed monthly discount rate\\
\bottomrule
\end{tabular}
\par\smallskip
\begin{minipage}{\textwidth}\footnotesize
\emph{Notes:} Time is measured in months and $K_0=1$. Targets from \citet{aifutures2026} are judgments in its worked example, used here as illustrative scenario inputs; assumed parameters are not estimated here. The pause target holds the compute clock fixed and retains baseline safety learning. Research capacity grows at $\gamma=\varepsilon g_K$. The fitted hazard density is $a_h e^{c_h F}$ on the full-speed path. Domestic hazards are normalized to $H(x)/N$ per lab. Complete values and alternative closure outcomes are recorded in the replication results.
\end{minipage}
\end{table}
\FloatBarrier

\subsection{Arrival Dates and the Value of a Month}

Slower research delays arrival, but less than in proportion. Let $T(x)$ be the arrival date at constant research share $x$, with safety fixed at $s_0$, and let $\tau(x)=-T'(x)>0$ be the months saved by a marginal increase in research. Both have closed forms for any $\phi$, derived in three steps.

\emph{Step 1: research capacity.} Dividing \eqref{eq:capability_ode} by $\phi(F)$ leaves only research capacity on the right side.

\emph{Step 2: the accumulated research.} Integrating from $F=0$ to $F=1$ at a constant share gives a left side that depends only on the technology. The right side is $(xK_0)^\varepsilon(e^{\gamma T(x)}-1)/\gamma$.

\emph{Step 3: the reference path.} The full-speed path gives the same left side with $x_0$ and $T_0$, so equating the two right sides gives
\begin{equation}
e^{\gamma T(x)}=1+\Bigl(\frac{x_0}{x}\Bigr)^{\varepsilon}\bigl(e^{\gamma T_0}-1\bigr),
\qquad
\tau(x)=\frac{1-e^{-\gamma T(x)}}{g_Kx}.
\label{eq:arrival}
\end{equation}
The two formulas in \eqref{eq:arrival} give the arrival date and the months saved by a marginal rise in research. The second follows by differentiating the first. Arrival moves less than in proportion because a slower lab reaches later capability levels with more compute. In the calibration, halving research from 50 to 25 percent delays arrival from 18 to \rrHalfT{} months. At the full-speed share $\tau(x_0)=\rrTauZero$, so one more percentage point of research brings arrival forward by \rrTauPoint{} months. Let $H(x)$ be cumulative hazard along that path. Define
\begin{equation}
V(x)\equiv\frac{H'(x)}{\tau(x)}.
\label{eq:V_def}
\end{equation}
We call $V$ the value of a month. Increasing research by $dx$ brings arrival forward by $\tau dx$ months and changes hazard by $H'dx$. So $V$ is the marginal hazard increase per month saved, or the hazard reduction per additional month when research is slowed. It measures cumulative hazard, not percentage points of catastrophe probability. We quote it in percent a month: a value of 1 percent means that one more month of delay lowers cumulative hazard by 0.01. 

The sign of $V$ depends on whether extra knowledge outweighs longer state exposure. Slowing research gives knowledge more time to build before each capability level, but spends more months at each level.

The general technology does not imply that $V$ is positive or increasing. With constant positive state hazard, no step hazard, a constant safety yield, and the baseline safety share, knowledge depends on calendar time alone and $V(x)=-he^{-S(T(x))}<0$. Slower development then adds exposure.

For the comparisons below, assume differentiable $T,H$, $\tau>0$, and strictly increasing $V$ on the relevant interval. These restrictions hold on the baseline numerical curve but not on every calibration.

No contest sensitivity makes 50 percent research a racing equilibrium when the takeoff period is shorter than about \lswSwitchTzero{} months. The same holds when the pause target, the safety-case probability used to fit the safety yield, is above about \lswSwitchPreg{} percent, or when the state-hazard weight, the share of full-speed hazard that is state hazard, is above about \lswSwitchOmega{} (Appendix~\ref{app:calibration}).

The safety yield $\sigma$ is what makes a month valuable. With constant $\sigma$ and the baseline safety share, knowledge on any path is $S=\sigma t$. Let $t_x(F)$ be the date at which the path at share $x$ reaches capability $F$, and $f_x(F)$ its speed $\dot F$ there. Then cumulative hazard is
\[
H(x)=\int_0^1\Bigl[\frac{h(F)}{f_x(F)}+k(F)\Bigr]e^{-\sigma t_x(F)}\,dF.
\]
Slowing research moves every date $t_x(F)$ later. Each extra month multiplies the hazard still ahead by $e^{-\sigma}$, but it also adds a month of state exposure at the current level. Without safety learning, $\sigma=0$, only the exposure remains. In the calibration $V$ is then negative at every research share, and neither racing labs nor a common rule would slow down.

A higher yield makes each month remove more of the hazard ahead, but it also leaves less hazard ahead to remove. So at each research share in the calibration, $V$ first rises and then falls as $\sigma$ grows. At the baseline yield, $V$ at the racing share is close to its peak, while at the slow shares of coordination it is well past it. A 10 percent higher yield lowers $V$ by \sgVDropCoop\% at $x_J$ but by only \sgVDropRace\% at 50\% research. Section~\ref{sec:numbers} shows what this does to the three regimes.

\section{Racing and Coordination}\label{sec:developers}

The domestic industry contains $N\ge2$ frontier labs. Earlier arrival reduces discounting, while a relative head start improves a lab's chance of winning a commercial prize.

This section has four parts. Proposition~\ref{prop:race} derives the racing wedge and splits it into the hazard externality and the contest part. Proposition~\ref{prop:labs} asks how the number of labs changes it. Proposition~\ref{prop:publish} shows that open source removes both parts of the wedge. Section~\ref{sec:unilateral} compares one lab's restraint with a common rule, and Section~\ref{sec:numbers} puts numbers on all four. Appendix~\ref{app:proofs} gives the proofs, and Appendix~\ref{app:notation} lists the notation.

\subsection{Market Structure and Payoffs}\label{sec:race_setup}

Each lab $n$ chooses a constant share $0<x_n\le1-s_0$, keeps safety at $s_0$, and assigns $1-x_n-s_0$ to customers. The game considers stationary shares rather than full dynamic strategies. Customer compute changes, but service revenue and consumer surplus do not enter the terminal-payoff objective. Each lab runs its own frontier on its own compute $K(t)$. So changing the number of labs below holds per-frontier compute and technology fixed; it does not divide a fixed industry stock.

Lab $n$ arrives at $T(x_n)$. Under the baseline normalization, its frontier contributes $H(x_n)/N$, so industry hazard is
\begin{equation}
H_{\text{ind}}=\frac1N\sum_{m=1}^N H(x_m).
\end{equation}
At a common share this equals $H(x)$ for every $N$. This normalization isolates the strategic effect of market structure. It does not estimate how adding an actual frontier changes physical risk; Section~\ref{sec:labs} also considers fixed per-frontier hazard.

A catastrophe from any frontier eliminates every lab's terminal stake. Conditional on survival over the industry's takeoff paths, lab $n$ has enterprise value $B>0$ and receives $\Pi>0$ if it wins the contest. Each lab discounts at its own arrival date, while its stake is contingent on survival across all modeled frontiers. This is a reduced-form terminal-stake objective, without intermediate revenue or lifetime welfare.

A logit contest allocates the prize \citep{mcfadden1974}:
\begin{equation}
p_n(\mathbf{x})=
\frac{e^{-\kappa T(x_n)}}{\sum_{m=1}^N e^{-\kappa T(x_m)}},
\label{eq:logit_contest}
\end{equation}
where $\kappa>0$ is the contest sensitivity, measured per month. This is the logistic, or difference-form, contest success function of \citet{hirshleifer1989}, with arrival dates, entered with a minus sign, in place of efforts; Hirshleifer calls $\kappa$ the mass effect. It sets how much the prize depends on arriving early. For any two labs, $p_n/p_m=e^{\kappa[T(x_m)-T(x_n)]}$, so each month of head start multiplies a lab's odds against a rival by $e^{\kappa}$. Only differences in arrival dates matter: if every lab arrived a month earlier, no lab's odds would change.

Among the contest success functions axiomatized by \citet{skaperdas1996}, only the logit has this property. Adding the same amount to every contestant's effort leaves the odds unchanged \citep{jia2013}.

Two limits bracket the contest. At $\kappa=0$ the prize is a lottery that ignores arrival dates, so speed earns no prize; this is the no-contest case of Proposition~\ref{prop:race}. As $\kappa$ grows, the first lab to arrive wins with near certainty, a winner-take-all race. Intermediate values describe a market in which arriving first helps but does not settle who leads, for example because customers can switch to a later model.

Section~\ref{sec:numbers} fits $\kappa=\dvKappa$ per month. A one-year head start then multiplies a lab's odds by \dvKappaYearOdds. A lab one year ahead of two level rivals wins the prize with probability \rrYearHeadWin\%, against one third when all three are level.

Risk attitudes shape effort and the chance of winning as well, and the sign of that effect depends on liability \citep{skaperdas1995}. The contest uses the deterministic path dates; the model does not separately integrate stochastic breakthrough times into hazard and discounting. So the logit contest is a reduced form: $p_n$ is a lab's probability of winning the commercial prize, not of arriving first.

At monthly discount rate $r>0$, lab $n$ maximizes
\begin{equation}
U_n(x_n;\mathbf{x}_{-n})=
\exp\left(-\frac1N\sum_{m=1}^N H(x_m)\right)
e^{-rT(x_n)}(\Pi p_n+B).
\label{eq:firm_payoff}
\end{equation}
Write $\pi=\Pi/(\Pi+NB)$ for the prize's share of terminal industry value. Survival gives a reason to slow; discounting and the contest give reasons to arrive sooner.

\subsection{Racing versus Coordination}\label{sec:race}

Compare symmetric racing with common-share coordination, the joint-profit optimum among common shares. It is the best common share, not the best industry allocation.

Three regimes organize the comparison. They differ in who chooses the research share and in what that choice maximizes.

\emph{Racing.} Each lab chooses its own share to maximize its own payoff~\eqref{eq:firm_payoff}, taking its rivals' shares as given. Arriving earlier raises its odds of winning the prize. The racing share $x_R$ is a symmetric Nash equilibrium: when every other lab uses $x_R$, no lab gains by choosing a different share.

\emph{Common-share coordination.} The industry sets one share $x$ for every lab to maximize the labs' total payoff. At a common share every lab arrives at $T(x)$, industry hazard is $H(x)$, and each lab wins the prize with probability $1/N$. Adding up the $N$ payoffs~\eqref{eq:firm_payoff} gives the joint payoff
\[
U_{\text{joint}}(x)=e^{-H(x)}e^{-rT(x)}(\Pi+NB).
\]
The prize goes to some lab whatever the share, so it no longer rewards speed. The coordinated share $x_J$ maximizes $U_{\text{joint}}$.

\emph{No contest.} Labs again choose separately, but the prize does not depend on speed. With $\kappa=0$ in~\eqref{eq:logit_contest}, each lab wins with probability $1/N$ whatever its arrival date, as in a lottery. Lab $n$'s payoff becomes
\[
U_n^E(x_n;\mathbf{x}_{-n})=
\exp\left(-\frac1N\sum_{m=1}^N H(x_m)\right)
e^{-rT(x_n)}\left(\frac{\Pi}{N}+B\right).
\]
The no-contest share $x_E$ is the symmetric equilibrium share of this game. Each lab still bears only one $N$th of the hazard its speed adds, so only this hazard externality separates $x_E$ from coordination. The subscripts stand for racing, joint, and externality.

The idea is simple. A lab bears only its share of the hazard its speed adds, but it keeps the whole timing benefit and the better odds of winning the prize. So its hurdle for slowing down, the racing hurdle, is higher than the industry's. The derivation takes three steps.

\emph{Step 1: the hazard cost.} A small rise $dx$ in one lab's research brings its arrival forward by $\tau dx$ months. It raises industry hazard by only $H'dx/N$, because that lab's frontier carries one $N$th of the hazard.

\emph{Step 2: the private gain.} Each month saved is worth $r$ in discounting and also improves the lab's odds of winning the prize. At a symmetric profile each lab wins with probability $1/N$, and each month saved adds $\kappa(N-1)\pi/N$ to the log payoff at the margin.

\emph{Step 3: the hurdle.} An interior equilibrium sets the hazard cost equal to the gain. Dividing by the months saved and multiplying by $N$ gives the hurdle $Nr+\kappa(N-1)\pi$. A common rule moves all frontiers together, so the industry weighs the full $H'dx$ against $r$ per month saved.

Table~\ref{tab:hurdles} sets the three regimes side by side, for one more month of delay. Each regime stops where the value of a month $V$ reaches its hurdle.

\begin{table}[htbp]
\centering\small
\caption{The Hurdle for One More Month of Delay in Each Regime}\label{tab:hurdles}
\begin{tabular}{@{}>{\raggedright\arraybackslash}p{.17\textwidth}>{\raggedright\arraybackslash}p{.25\textwidth}>{\raggedright\arraybackslash}p{.25\textwidth}>{\raggedright\arraybackslash}p{.25\textwidth}@{}}
\toprule
 & Common rule & Racing lab & Racing lab, no contest\\
\midrule
Hazard counted & All of it, at every lab, $H'$ & Its own share, $H'/N$ & Its own share, $H'/N$\\
Gain per month saved & $r$, since the prize goes to some lab whatever the common share & $r+\kappa(N-1)\pi/N$, discounting plus better odds of the prize & $r$\\
Accepts the month if & $V\ge r$ & $V\ge Nr+\kappa(N-1)\pi$ & $V\ge Nr$\\
Calibration, percent a month & \dvRMonthPct & \dvVRacePct & \rrNoContestHurdle\\
\bottomrule
\end{tabular}
\end{table}

The racing wedge is the gap between the first two hurdles.

\begin{proposition}[The Racing Wedge]\label{prop:race}
Any interior symmetric Nash equilibrium share $x_R$ satisfies
\begin{equation}
V(x_R)=Nr+\kappa(N-1)\pi.
\label{eq:race_foc}
\end{equation}
An interior coordinated share $x_J$, the joint-profit optimum among common shares, satisfies
\begin{equation}
V(x_J)=r.
\label{eq:coord_foc}
\end{equation}
With no speed-dependent contest reward ($\kappa=0$), an interior symmetric equilibrium share $x_E$ satisfies
\begin{equation}
V(x_E)=Nr.
\label{eq:externality_foc}
\end{equation}
If these solutions exist in the interior and $V$ is strictly increasing, then $x_J<x_E<x_R$ and $T(x_R)<T(x_E)<T(x_J)$. At an upper share constraint a boundary condition replaces the first-order equality.
\end{proposition}
The proof is in Appendix~\ref{app:proofs}.

Racing is faster than coordination because two forces raise a racing lab's hurdle above the discount rate $r$. First, each lab counts only one $N$th of the hazard its speed adds, which turns $r$ into $Nr$; this hazard externality adds $(N-1)r$. Second, a private head start takes winning odds from the other labs even though the industry prize is fixed. That business stealing adds $\kappa(N-1)\pi$, the contest part of the wedge. So discounting $r$, the hazard externality $(N-1)r$, and the contest part make up the racing hurdle. Without the contest it is $Nr$, the no-contest hurdle.

In the calibration of Table~\ref{tab:calibration}, $N=\dvN$, $r=\dvRMonthPct\%$ per month, $\kappa=\dvKappa$ and $\pi=0.\dvPiShare$. The racing hurdle of \dvVRacePct\% a month is discounting at \dvRMonthPct\%, plus the hazard externality at \rrExternality\%, plus the contest part at \rrContestPart\%. It is \dvVRaceOverR{} times the hurdle at which coordination stops.

\subsection{Market Structure and Fragmentation}\label{sec:labs}

More labs raise each lab's hurdle, through both of its parts. Keep the technology, $\kappa$, $B$, and $\Pi$ fixed when changing $N$.

\emph{Step 1: the hazard share.} Each frontier carries one $N$th of the industry hazard, so an added lab lowers the share of its own hazard that each lab counts. The no-contest hurdle $Nr$ rises by $r$.

\emph{Step 2: the contest.} An added lab is one more rival whose prize can be taken, so $N-1$ rises by one. The prize share $\pi=\Pi/(\Pi+NB)$ falls, but the product $(N-1)\pi$ still rises, as Appendix~\ref{app:proofs} shows.

\emph{Step 3: research shares.} With $V$ increasing, a higher hurdle means a higher racing share, until the upper constraint binds.

\begin{proposition}[Market Fragmentation and Racing]\label{prop:labs}
Under the baseline hazard normalization:
\begin{enumerate}
\item[(a)] The symmetric hurdle $Nr+\kappa(N-1)\Pi/(\Pi+NB)$ strictly increases in $N$.
\item[(b)] If $V$ is strictly increasing and the compared symmetric equilibria exist, their research shares are weakly increasing in $N$. The ordering is strict when both are interior and can be flat at the upper constraint.
\end{enumerate}
\end{proposition}
The proof is in Appendix~\ref{app:proofs}.

For example, at $\Pi=3B$ and the calibrated $\kappa$, the hurdle is \rrHurdleTwo\% per month for two labs, \dvVRacePct\% for three, and \rrHurdleFour\% for four. Research rises from \dvLabsTwoX\% to 50\% to the 95\% upper constraint.

\citet{bdmdp2026} find that fragmentation shifts a fixed pool of industry resources toward speed. Here the experiment differs, since each lab keeps its own compute and technology as labs are added. Each lab's hurdle rises with the number of labs, and with a fixed per-frontier hazard research shares can fall as labs are added.

Hazard can instead be fixed per frontier. Each lab then carries $H(x)/N_0$, where $N_0=3$. The survival factor uses $\sum_mH(x_m)/N_0$. Only the hazard weight in the first-order condition changes: each lab now counts $1/N_0$ of its own hazard. Multiplying that condition by $N_0/\tau$ gives the interior symmetric condition
\begin{equation}
V(x_R)=N_0r+
N_0\kappa\left(1-\frac1N\right)\frac{\Pi}{\Pi+NB}.
\label{eq:fixed_frontier_foc}
\end{equation}
The no-contest hurdle $N_0r$ is constant in $N$. With $\Pi=3B$, the contest part peaks at $N=3$ and falls beyond it, so research need not rise with fragmentation. Aggregate hazard is $NH(x_R(N))/N_0$, so more frontiers add risk, but equilibrium shares can fall. Both effects must be checked. In the calibrated cases $N=1,\ldots,5$, aggregate risk increases. This is a numerical result rather than an unrestricted monotonicity theorem.

\subsection{Open Source and Transparent Compute}\label{sec:publish}

So far each lab has its own frontier, and its findings stay private. Now suppose every lab publishes its capability findings at once. We call this open source, in its ideal form. All domestic labs then work from one shared frontier, and each lab's research moves it for everyone. Monitoring how much compute a lab spends on safety is a different policy, because it makes no finding public.

Normalize published input as $z(\mathbf{x})=N^{-1}\sum_nx_n$, so symmetric contributions $x_n=x$ reproduce $z=x$. Assume all labs arrive at $T(z)$, industry hazard is $H(z)$, and safety knowledge is the same along the common path. Retain the terminal objective with no separate contribution cost beyond its effect on that path.

\emph{Step 1: the contest disappears.} A lab's contribution moves the common frontier, so it moves every lab's arrival together. Relative arrival does not change, and each lab wins the prize with probability $1/N$.

\emph{Step 2: the hazard externality disappears.} The same contribution changes the shared hazard $H(z)$ and every lab's timing in the same proportion, $1/N$. So each lab weighs the whole industry hazard against its own discounting, as a common rule does.

\begin{proposition}[Open Source Aligns Contribution Incentives]\label{prop:publish}
Under open source, $p_n=1/N$ independently of unilateral contributions. Any interior equilibrium in effective input satisfies
\begin{equation}
V(z)=r.
\end{equation}
This is a condition on the aggregate input. It does not uniquely determine each contribution or automatically cover private research outside the shared frontier.
\end{proposition}
The proof is in Appendix~\ref{app:proofs}.

In words, open source removes both parts of the wedge. No lab can get ahead of the others, because every finding reaches all of them at once. And each lab's research raises the one shared hazard, so each lab counts all of it. Removing the contest alone, with separate frontiers, leaves $V(x_E)=Nr$. For the $N=3$ calibration, removing only the contest leaves labs at \dvXNoContest\% research, whereas open source lowers the effective input to the coordinated share of \dvXCoop\% if that solution is interior. Different individual contributions can still sum to the same average input $z$. Real releases fall short of this ideal in three ways. A release that comes with a lag gives the publishing lab a head start for that long, so part of the contest returns. Research kept private, outside the shared frontier, gives labs a private reason to speed up again. And published findings also reach foreign rivals, a channel this paper leaves out.

Capability work remains useful where $V(z)<r$. Publishing capability findings, publishing safety findings and enforcing a safety allocation have different effects. The proposition is a benchmark for complete and immediate publication, not a claim that transparency alone prevents diversion or enforces itself. Publication limited to domestic labs would also need a limit on foreign access.

\subsection{Unilateral Restraint versus Common Rules}\label{sec:unilateral}

Suppose Lab 1 commits to $\bar x<x_R$, while rivals choose best responses. A unilateral commitment does not bind the rivals' shares, and whether they slow at all depends on the prize structure. Restraint changes their odds of winning the prize. Near a symmetric profile, rivals slow when the prize is large relative to the stake, $\Pi/B>N(N-2)$, and speed up when it is smaller (Appendix~\ref{app:cross}). The baseline, $N=3$ and $\Pi/B=3$, sits exactly at the boundary, so the local response is zero. Larger restraint still makes rivals slow, as Section~\ref{sec:numbers} shows.

The numerical comparison asks what fraction of the same restriction's industry-wide risk reduction one restrained lab achieves, allowing rivals to adjust.

A voluntary pledge to adopt a common share would not hold. At the coordinated share each lab would gain by speeding up again, because the value of a month falls short of the racing hurdle (Proposition~\ref{prop:race}). Section~\ref{sec:numbers} measures what one lab's restraint achieves.

\section{Quantitative Results}\label{sec:numbers}

\subsection{Racing and Coordination}

The calibration imposes the racing share and fits the contest to it. Set $N=\dvN$, $r=\dvRMonthPct\%$ per month ($0.03/12$), and $\pi=0.\dvPiShare$, so $\Pi=\dvPrizeOverB B$. Fitting $\kappa$ so the symmetric condition \eqref{eq:race_foc} reproduces the 50\% full-speed share gives $\kappa=\dvKappa$ per month. A one-year head start then multiplies a lab's odds relative to a rival by $\dvKappaYearOdds$. The racing share pins down only the product $\kappa\pi$, so the prize share is set first (Appendix~\ref{app:calibration}).

\begin{table}[htbp]
\centering
\caption{Racing Labs and Common-Share Coordination}
\label{tab:race}
\small
\setlength{\tabcolsep}{2.5pt}
\begin{tabular}{@{}lccc@{}}
\toprule
& Research share & Takeoff period & Catastrophe\\
Specification & (\% of compute) & (months) & probability (\%)\\
\midrule
Three racing labs (symmetric Nash equilibrium) & 50.0 & 18.0 & \dvRiskRace\\
Three labs, no speed-dependent contest ($\kappa=0$) & \dvXNoContest & \dvTNoContest & \dvRiskNoContest\\
Common-share coordination & \dvXCoop & \dvTCoop & \dvRiskCoop\\
Common-share coordination, maximum 12-month delay & \dvLeadCoopX & \dvLeadCoopT & \dvLeadCoopRisk\\
\midrule
Two racing labs ($N=2$) & \dvLabsTwoX & \dvLabsTwoT & \dvLabsTwoRisk\\
Four or five racing labs ($N=4,5$) & \dvLabsFourX & \dvLabsFourT & \dvLabsFourRisk\\
Contest half as decisive ($\kappa/2$) & \dvKappaHalfX & \dvKappaHalfT & \dvKappaHalfRisk\\
Contest twice as decisive ($2\kappa$) & \dvKappaDoubleX & \dvKappaDoubleT & \dvKappaDoubleRisk\\
\bottomrule
\end{tabular}
\par\smallskip
\begin{minipage}{\textwidth}
\footnotesize
\emph{Notes:} Scenario calculations with every lab's safety share held at 5\%, so research cannot exceed 95\%. Coordination optimizes survival-weighted discounted terminal industry value among common shares; its interior condition is $V=r$. The probability is $1-e^{-H_{\rm ind}}$.
\end{minipage}
\end{table}

In Table~\ref{tab:race}, symmetric racing uses 50\% research, takes 18 months, and carries a $\dvRiskRace\%$ catastrophe probability. Common-share coordination lowers research to $\dvXCoop\%$, lengthens the takeoff period to $\dvTCoop$ months, and lowers the catastrophe probability to $\dvRiskCoop\%$. If the common share may delay arrival by at most 12 months, the best one uses \dvLeadCoopX\% research, takes \dvLeadCoopT{} months, and gives \dvLeadCoopRisk\%.

\subsection{Decomposing the Racing Wedge and the Probability Gap}

Removing the speed-dependent contest while retaining the hazard externality gives $\dvXNoContest\%$ research and a catastrophe probability of $\dvRiskNoContest\%$. The contest accounts for about \rrMarginalShare\% of the marginal wedge, the share
\[
\frac{\kappa(N-1)\pi}{Nr+\kappa(N-1)\pi-r}.
\]
The numerator is the contest part; the denominator is the wedge, the gap between the racing and coordination hurdles. The contest accounts for about \rrProbShare\% of the probability-point gap, the share
\[
\frac{P_R-P_E}{P_R-P_J}.
\]
The numerator is the risk removed by eliminating the contest alone; the denominator is the risk removed by coordination. In the baseline, the contest removes \rrGapContest{} of the \rrGapCoord{} percentage points that coordination removes. The first share decomposes a marginal condition; the second is a sequential counterfactual through the no-contest equilibrium. Neither is an estimated industry-wide percentage.

The marginal share is set largely by the calibration target. Fitting $\kappa$ so that the race lands on the imposed 50 percent share gives $\kappa(N-1)\pi=V(x_R)-Nr$, so the marginal share equals $[V(x_R)-Nr]/[V(x_R)-r]$, fixed by the racing target, $N$ and $r$. Changing the prize share while refitting $\kappa$ to the same target cannot test it. What the number shows is that discounting and the hazard externality together, $Nr=\rrNoContestHurdle\%$ a month, cannot explain racing where the value of a month is \dvVRacePct\%. Holding $\kappa$ fixed and lowering the prize share does move the result. A smaller prize lowers the probability-gap attribution to about \rrSensPrizeShare\% (Appendix~\ref{app:calibration}).

The probability-gap share also depends on the order of the counterfactual. Remove the hazard externality first instead. Each lab still competes for the prize, but it counts in full the hazard that its own research adds, instead of one $N$th of it. Its payoff becomes
\[
U_n^A(x_n;\mathbf{x}_{-n})=
\exp\left(-H(x_n)-\frac1N\sum_{m\ne n}H(x_m)\right)
e^{-rT(x_n)}(\Pi p_n+B).
\]
In Step~1 of Section~\ref{sec:race} the hazard cost of speed is then $H'$ rather than $H'/N$, while the gain per month saved is unchanged. So labs stop where $V=r+\kappa(N-1)\pi/N=\dcHurdleNoExt\%$ a month. Racing then uses $x_A=\dcXNoExt\%$ research with risk $P_A=\dcRiskNoExt\%$, and the contest accounts for the share $(P_A-P_J)/(P_R-P_J)$, about \dcContestSecond\% of the gap. The average over the two orders is about \dcContestShapley\%.

Each force alone supports little research. Without the contest, labs use \dvXNoContest\%, and without the hazard externality \dcXNoExt\%. Racing at 50 percent needs both, because counting only one $N$th of the hazard multiplies every private gain from speed by $N$, the contest's included.

\begin{figure}[htbp]
\centering
\includegraphics[width=\textwidth]{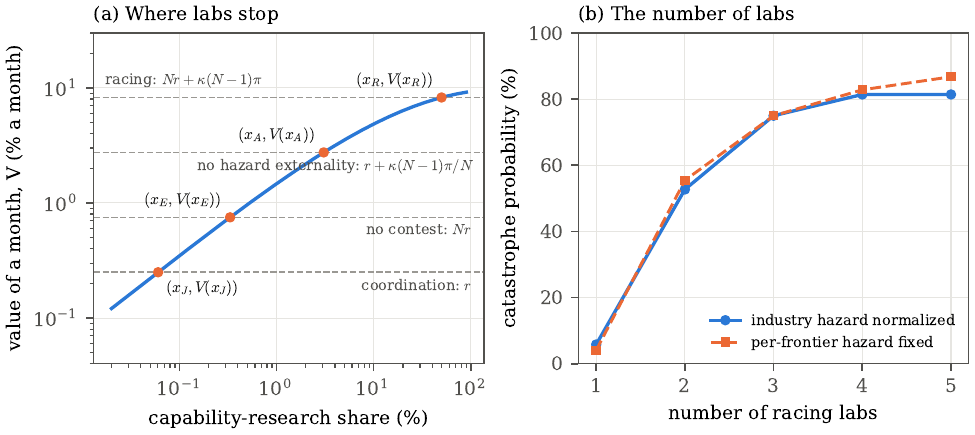}
\caption{(a) The value of a month, $V$, against research share, with the hurdle for each combination of forces. An interior symmetric race stops where $V=Nr+\kappa(N-1)\pi$, at $x_R$. Without the hazard externality labs stop where $V=r+\kappa(N-1)\pi/N$, at $x_A$; without the contest, where $V=Nr$, at $x_E$. Common-share coordination stops where $V=r$, at $x_J$. (b) Calibrated catastrophe probability against the number of labs. The solid line normalizes industry hazard at a common share; the dashed line holds each frontier's hazard fixed.}
\label{fig:developers}
\end{figure}

Figure~\ref{fig:developers}a shows the four hurdles on the increasing $V$ curve. From the top, they are racing ($x_R$), no hazard externality ($x_A$), no contest ($x_E$), and coordination ($x_J$). Figure~\ref{fig:developers}b plots the catastrophe probability against the number of labs. Along the solid line, two labs race at $\dvLabsTwoX\%$ research, three at 50\%, and four or five at the same 95\% upper constraint. The dashed line holds each frontier's hazard fixed; aggregate risk still rises with the number of labs in the calibrated cases.

\subsection{Safety Effort and Safety Yield}

Safety effort is fixed in the game, and the restriction does real work. Under the same terminal payoff, customer compute has no value, so a lab free to choose safety would move it into safety. Table~\ref{tab:safety} instead imposes higher safety shares on every lab, holding the hazard primitives and $\kappa$ fixed. With 25 percent safety, racing labs hit the upper constraint of \lssRaceXTwentyFive\% research, against 50\% at the baseline, and racing risk falls to \lssRacePTwentyFive\%. Labs offset part of the extra safety by racing faster. The contest's share of the probability gap stays near \lssShareTwentyFive\%. Section~\ref{sec:accord} applies this result to the safety audits of the September 2026 accord.

The safety yield works through the same channel, since a higher safety share multiplies the knowledge rate by $y(s)$. Table~\ref{tab:sigma} in Appendix~\ref{app:calibration} scales $\sigma$, holding the hazard primitives, $\kappa$, $\pi$ and $r$ fixed. A higher yield lowers risk under every regime, and labs again offset part of the gain by racing faster. At 1.25 times the baseline yield, full-speed risk falls from \dvRiskRace{} to \sgFullPOneQ\%. Racing labs speed up from 50 to \sgRaceXOneQ\% research, so racing risk falls only to \sgRacePOneQ\%. Across the higher yields in the table, the faster race gives back \sgGivenBackMin--\sgGivenBackMax\% of the safety gain.

A common rule speeds up too, from \sgCoopXBase\% to \sgCoopXOneQ\% research, because $V$ falls most at slow shares, while its risk falls from \sgCoopPBase\% to \sgCoopPOneQ\%. The contest's share of the gap stays at \sgShareMin--\sgShareMax\%. So the yield changes how risky every regime is, while the contest explains a similar share of the gap.

A lower yield pushes the race to its upper constraint. At three quarters of the baseline, $V$ peaks at \sgPeakThreeQ\% research without reaching the racing hurdle, so racing labs use 95\% research and carry \sgRacePThreeQ\% risk. At half the baseline the tested grid has no symmetric equilibrium. So the assumption that $V$ rises with research share needs a yield not far below the baseline.

\begin{table}[htbp]\centering\small
\caption{Racing at Different Imposed Safety Shares}\label{tab:safety}
\setlength{\tabcolsep}{4pt}
\begin{tabular}{@{}ccccccc@{}}\toprule
Safety share & \multicolumn{2}{c}{Racing labs} & \multicolumn{2}{c}{No contest ($\kappa=0$)} & Coordination & Contest share\\ \cmidrule(lr){2-3}\cmidrule(lr){4-5}
(\%) & Research (\%) & Risk (\%) & Research (\%) & Risk (\%) & risk (\%) & of the gap (\%)\\ \midrule
5 & 50.0 & 75.0 & 0.33 & 11.8 & 5.8 & 91\\
10 & 58.9 & 67.5 & 0.81 & 10.2 & 5.0 & 92\\
25 & 75.0$^{\dagger}$ & 53.6 & 2.71 & 8.2 & 4.1 & 92\\
\bottomrule\end{tabular}
\par\smallskip\begin{minipage}{\textwidth}\footnotesize
\emph{Notes:} Every lab is held at the stated safety share; customers receive the rest. Research cannot exceed $1-s$. The physical hazard primitives and the fitted contest sensitivity $\kappa$ are those of the baseline (5\% safety) and are not refitted, so the race need not stay at 50\% research. $^{\dagger}$The race sits at the upper constraint. The last column is $(P_R-P_E)/(P_R-P_J)$. Section~\ref{sec:accord} applies the table to the safety audits of the 2026 White House Accord on Super Intelligence.
\end{minipage}\end{table}

\subsection{Unilateral Restraint}

One lab's restraint achieves much less than a common rule. Table~\ref{tab:uni} holds Lab 1 at each reported share and computes an equilibrium among its rivals.

\begin{table}[htbp]
\centering
\caption{One Lab's Unilateral Restraint versus a Common Rule}
\label{tab:uni}
\small
\setlength{\tabcolsep}{6pt}
\begin{tabular}{@{}lcccc@{}}
\toprule
& & Catastrophe & Catastrophe &\\
Restrained lab's & Rivals' response & probability, & probability, & Share of common\\
research share (\%) & research share (\%) & unilateral (\%) & common rule (\%) & rule's gain (\%)\\
\midrule
Coordinated share ($\dvXCoop\%$) & \dvUniCoopOthers & \dvUniCoopRisk & \dvUniCoopCommon & \dvUniCoopShare\\
5.0\% & \dvUniFiveOthers & \dvUniFiveRisk & \dvUniFiveCommon & \dvUniFiveShare\\
10.0\% & \dvUniTenOthers & \dvUniTenRisk & \dvUniTenCommon & \dvUniTenShare\\
\bottomrule
\end{tabular}
\par\smallskip
\begin{minipage}{\textwidth}
\footnotesize
\emph{Notes:} The baseline has all labs at 50\% research and $\dvRiskRace\%$ risk. Column 2 gives the selected symmetric equilibrium of the remaining labs conditional on Lab 1's share. Column 5 is $(P_R-P_{\rm unilateral})/(P_R-P_{\rm common})\times100$. These are finite scenario changes, not local complementarity results.
\end{minipage}
\end{table}

When Lab 1 slows to the coordinated share ($\dvXCoop\%$), its rivals slow only from 50\% to $\dvUniCoopOthers\%$. Industry risk falls to $\dvUniCoopRisk\%$, achieving $\dvUniCoopShare\%$ of the reduction from a common rule at the same share. Restraint to 5\% or 10\% research likewise achieves much less than a common rule.

\subsection{The Hazard Level}\label{sec:hazard_level}

The baseline takes the AI Futures Project's 75 percent probability of catastrophe at full speed \citep{aifutures2026}. \citet{greenblatt2025} gives the same 75 percent for a takeoff period with almost no safety effort. We also report 25 percent, the probability \citet{samotsvety2022} give for misaligned AI takeover by 2100. Because the model conditions on a full-speed takeoff period, its probability should sit above unconditional estimates like that one, so 25 percent is a cautious lower case. Surveys of AI researchers report medians of 5 to 10 percent \citep{grace2025}.

A lower hazard level keeps racing faster than a common rule and shrinks the racing wedge. Table~\ref{tab:hazard_level} refits the calibration at 25 percent. A racing lab then accepts a month of delay only for a hazard cut of \hzrVLo{} percent a month, \hzrMultLo{} times the common rule's hurdle, against \hzrVHi{} percent and \hzrMultHi{} times at 75 percent. The gap in catastrophe probability between racing and coordination falls from \hzrGapHi{} to \hzrGapLo{} points.

The contest explains less of that gap. Removing the contest first closes \hzrFirstLo{} percent of it, against \hzrFirstHi{} percent at 75 percent, and the average over the two orders falls below one half, to \hzrShapLo{} percent. The one-sided counterfactuals also swap order. Without the contest labs would use \hzrXNoContestLo{} percent research, and without the hazard externality \hzrXNoExtLo{} percent; at 75 percent the two shares are \hzrXNoContestHi{} and \hzrXNoExtHi{} percent. So the attribution to the contest holds at the upper end of the range, and at the lower end the hazard externality explains the larger part on average. Appendix~\ref{app:calibration} shows that below a full-speed probability of about \hzFloor{} percent no contest sensitivity supports the 50 percent racing share.

\begin{table}[htbp]
\centering\small
\caption{The Hazard Level: The Baseline and a Lower Case}\label{tab:hazard_level}
\begin{tabular}{@{}lrr@{}}
\toprule
& \multicolumn{2}{c}{Full-speed catastrophe probability}\\
\cmidrule(l){2-3}
& 75\% (baseline) & 25\%\\
\midrule
Racing catastrophe probability (\%) & \hzrRaceHi & \hzrRaceLo\\
Common-rule catastrophe probability (\%) & \hzrCoopHi & \hzrCoopLo\\
Common-rule research share (\%) & \hzrXCoopHi & \hzrXCoopLo\\
Value of a month at the racing share (\% a month) & \hzrVHi & \hzrVLo\\
\quad as a multiple of the common rule's hurdle & \hzrMultHi & \hzrMultLo\\
Research share without the contest (\%) & \hzrXNoContestHi & \hzrXNoContestLo\\
Research share without the hazard externality (\%) & \hzrXNoExtHi & \hzrXNoExtLo\\
Contest's share of the gap, removed first (\%) & \hzrFirstHi & \hzrFirstLo\\
Contest's share of the gap, removed second (\%) & \hzrSecondHi & \hzrSecondLo\\
Contest's share of the gap, average of the two orders (\%) & \hzrShapHi & \hzrShapLo\\
\bottomrule
\end{tabular}
\par\smallskip\begin{minipage}{\textwidth}\footnotesize
\emph{Notes:} Three labs, racing at 50\% research in both columns. Each column refits the safety yield, the hazard scale and the contest sensitivity to its full-speed probability, and sets the safety-case regime's probability so that it cuts cumulative hazard in the same proportion (\hzrRegHi\% at 75\%, \hzrRegLo\% at 25\%). The common rule's hurdle is the discount rate, $\dvRMonthPct\%$ a month.
\end{minipage}
\end{table}

\section{Application: The White House Accord on Super Intelligence}\label{sec:accord}

On September 29, 2026, the President and the heads of six AI and chip companies signed the White House Accord on Super Intelligence \citep{whitehouse2026}. The companies are Google, Anthropic, Meta, OpenAI, xAI and Nvidia. It asks each company to put in place four layers of controls and audits. The first is internal controls that monitor the capabilities and alignment of its models, in areas such as cybersecurity, biosecurity and chemical threats. The second is an internal team that makes sure the controls work. The third is an independent external auditor or evaluator, and the fourth an independent board committee that oversees the results. The companies will also meet regularly to set standards and best practices for safety.

The accord is voluntary. The companies say they are ``committed to doing this'' but leave law for later. Their text notes that ``it may make sense to codify these steps into laws or regulations'' over time. It sets no penalties, and it says nothing about how fast labs develop. The common rule of this paper differs on both counts: it sets every lab's research share, and it binds every lab.

In the model, the accord will not slow the race. It leaves each lab's research share free, so each lab still faces the racing hurdle $Nr+\kappa(N-1)\pi$ of Proposition~\ref{prop:race}. Racing labs still put 50 percent of compute into research and stop where the value of a month is \dvVRacePct{} percent a month, against $r$ under coordination. A voluntary pledge on speed would not slow the race either, because it would not hold (Section~\ref{sec:unilateral}). If only one lab kept such a pledge, its rivals would slow from 50 to \dvUniCoopOthers{} percent. Risk would fall to \dvUniCoopRisk{} percent, \dvUniCoopShare{} percent of the reduction a common rule achieves (Table~\ref{tab:uni}).

The accord's audits act on safety effort, not on speed, and the model bounds what they can do through that channel. Table~\ref{tab:safety} raises every lab's safety share. At 25 percent safety, racing labs move to \lssRaceXTwentyFive{} percent research, and racing risk falls only from \lssRacePFive{} to \lssRacePTwentyFive{} percent. The same safety share under a common rule gives \lssCoopPTwentyFive{} percent. If audits instead made safety research more productive, Table~\ref{tab:sigma} shows that labs race faster as risk falls, giving back \sgGivenBackMin--\sgGivenBackMax\% of the gain. In both tables the contest still accounts for over ninety percent of the gap between racing and coordination. So the accord can lower risk, but it leaves each lab's gain from speeding up essentially unchanged.

The model does not measure the monitoring and oversight processes themselves, and the accord does not require audit results to be made public. So the model says nothing about how well those processes work. What it does say is what drives labs to speed up and what removes that drive. In the model, two things slow the race. One is a common rule on research shares that binds every lab. The other is open source (Proposition~\ref{prop:publish}). Codifying the accord's four layers into law, as its text contemplates, would make them binding, but it would still leave speed to each lab. Section~\ref{sec:conclusion} returns to the competition-policy questions that a common rule raises.

\section{Discussion and Conclusion}\label{sec:conclusion}\label{sec:antitrust}\label{sec:scope}

Commercial leadership prizes accelerate frontier development relative to common-share coordination. Each lab counts only its share of the shared hazard while keeping the full timing gain and the better odds of winning the prize. In the three-lab calibration, removing the contest first closes over ninety percent of the gap between racing and coordination, and removing it after the hazard externality closes \dcContestSecond{} percent, or \dcContestShapley{} percent on average over the two orders. These shares are for a full-speed probability of 75 percent; at 25 percent they are \hzrFirstLo{}, \hzrSecondLo{} and \hzrShapLo{} percent (Section~\ref{sec:hazard_level}). Neither force alone explains racing at 50 percent.

A common rule achieves a much larger risk reduction than unilateral restraint. Open source also aligns each lab's incentive with coordination, when every finding reaches every lab at once.

The September 2026 accord leaves research shares free, so in the model it will not slow the race. Its audits can lower risk only through safety effort, within the limit of Table~\ref{tab:safety} (Section~\ref{sec:accord}).

Implementing a common rule raises competition-policy and enforcement questions. A common rule shifts compute from capability research to customer inference at a given capability level while delaying capability improvements over calendar time. Because the terminal-payoff game has no prices or demand, this shift does not by itself establish higher consumer surplus. Department of Justice guidance under the National Cooperative Research and Production Act evaluates qualifying joint ventures and standards-development activities under a rule of reason \citep{dojncrpa}. The model supports distinguishing research-pacing and safety standards from agreements on prices or output, without proving an antitrust exemption harmless. The September 2026 accord shows the same tension. Its signatories will ``meet regularly to establish standards'' \citep{whitehouse2026}, which may raise antitrust concerns \citep{crowell2026}, yet its commitments concern safety controls, not speed. Equal compliance also preserves relative arrival, so a secret unilateral advance still improves a deviator's odds. How detection, response time, and monitoring costs sustain restraint is the subject of \citet{gan2026agreements}.

Four limits qualify the results. First, the game restricts research shares to be stationary, fixes safety effort (varied in Table~\ref{tab:safety}), and leaves customer services out of the terminal payoff. It compares symmetric equilibria with common-share coordination under increasing $V$. Dynamic strategies, observable leads, asymmetric allocations, and service revenues could alter both equilibrium speed and welfare \citep{fudenberg1982,harris1987}. Consumer surplus and catastrophe losses outside the labs' terminal stakes would also change the welfare benchmark. Second, compute grows exogenously, and domestic capability sharing interacts with foreign access. Third, safety learning has share curvature but does not scale with total compute. Fourth, the calibration uses illustrative targets rather than identified causal estimates, and numerical best-response checks verify the reported equilibria without proving global uniqueness.

\appendix
\section{Calibration and Numerical Implementation}\label{app:calibration}

\subsection{Baseline Hazard and Pause Target}

Table~\ref{tab:calibration} separates imposed targets, assumed inputs, fitted parameters, and model implications. This appendix makes the baseline hazard construction and pause target explicit. These calibrations define an illustrative scenario, rather than identify a causal relationship from observed AI outcomes.

Normalize $K_0=1$. The calibration sets $\phi(F)=\theta e^{\rho F}$ with $\rho=\ln10$, and $y(s)=\psi+(1-\psi)(s/s_0)^\alpha$ with $\psi=\alpha=1/2$. At the baseline research share $x_0$, the initial research coefficient and the calendar date at capability $F$ are
\begin{align}
b_0&=\frac{\gamma(1-e^{-\rho})}{\rho(e^{\gamma T_0}-1)},
&\theta&=\frac{b_0}{x_0^\varepsilon},\\
t_0(F)&=\frac1\gamma\ln\left[1+\frac{e^{\gamma T_0}-1}{1-e^{-\rho}}
(1-e^{-\rho F})\right].
\end{align}
Here $b_0$ is the research speed at the start of the full-speed path, and $\theta$ converts it into productivity at the Automated Coder for $K_0=1$. The date $t_0(F)$ is when the full-speed path reaches capability $F$. The full-speed velocity at that capability is $f_0(F)=b_0e^{\gamma t_0(F)+\rho F}$, and baseline safety knowledge is $S_0(F)=\sigma t_0(F)$, starting from zero at $F=0$.

One target fixes the shape of the hazard. Let $q$ be the fraction of full-speed cumulative hazard occurring in the second half of the \emph{capability interval}, $F\in[1/2,1]$. It is not the fraction in the last half of calendar time. With hazard density $a_h e^{c_hF}$ per unit of capability,
\begin{equation}
c_h=2\ln\frac{q}{1-q},\qquad
a_h=\frac{-\ln(1-P_{\rm full})}{\int_0^1 e^{c_hF}\,dF}.
\end{equation}
The shape $c_h$ sets how fast the hazard density rises with capability; with $q=0.75$ it is $2\ln3\approx2.20$. The scale $a_h$ makes full-speed cumulative hazard equal $-\ln(1-P_{\rm full})=\ln4$. For a state-hazard share $\omega$, define the zero-knowledge primitives
\begin{equation}
h(F)=\omega a_h e^{c_hF} f_0(F)e^{S_0(F)},\qquad
k(F)=(1-\omega)a_h e^{c_hF}e^{S_0(F)}.
\end{equation}
Along the reference path, multiplying by $e^{-S_0}$ and using $dt=dF/f_0$ gives $\lambda_0dt=a_h e^{c_hF}dF$. State and step terms contribute fractions $\omega$ and $1-\omega$ of full-speed hazard. Holding these primitives fixed on alternative paths allows accumulated knowledge and the pace of progress to change risk.

The imposed pause-risk target suspends capability progress for $L_{\rm cal}$ months at $F=0$, keeps baseline safety research, and stops the compute clock. It then restarts the same reference capability path with the accumulated knowledge. With a constant safety yield, $\sigma$ solves
\begin{equation}
-\ln(1-P_{\rm reg})
=\frac{h(0)}{\sigma}(1-e^{-\sigma L_{\rm cal}})
+e^{-\sigma L_{\rm cal}}[-\ln(1-P_{\rm full})].
\end{equation}
The first term is state exposure during the pause; the second is subsequent hazard reduced by the additional safety stock. The 40\% target is the AI Futures Project's judgment for a safety-case regime \citep{aifutures2026}; assigning it to this specific stopped-clock experiment is a modeling assumption.

\subsection{Fitting the Contest}

At fixed $N$ and physical technology, one imposed interior racing share identifies only the product
\[
\kappa\pi=\frac{V(x_R)-Nr}{N-1}.
\]
The racing target fixes how much the contest must add, $V(x_R)-Nr$, and spreads it over the $N-1$ rivals. It cannot say how much comes from $\kappa$ and how much from $\pi$. Separately choosing a prize share is necessary to fit contest sensitivity. Table~\ref{tab:domestic_sensitivity} varies the imposed racing share, the state-hazard weight, and the prize. Refitting a 25\% racing target lowers racing risk to \rrSensTargetLow\%, while a 75\% target raises it to \rrSensTargetHigh\%. Holding $\kappa$ fixed and reducing $\pi$ to 0.25 produces \rrSensPrizeRisk\% risk and an \rrSensPrizeShare\% contest share of the coordination probability gap. So the baseline's over-ninety-percent attribution is not universal.

Changing the state-hazard weight refits $\sigma$ to retain the two hazard targets and then refits $\kappa$ to the 50\% racing target. With $\omega=0.75$, $V$ rises over the coordinated-to-racing comparison interval but peaks near $x=\rrSensPeak$ and declines thereafter. So increasing $V$ fails over the entire feasible interval in this case. The imposed racing share and no-contest share pass direct numerical best-response checks; a first-order condition or a global monotonicity claim alone would be insufficient. These exercises vary hazard composition and fitted safety yield while retaining the baseline share curvature and compute elasticity; they do not exhaust alternative safety technologies.

\begin{table}[htbp]
\centering\small
\caption{Domestic sensitivity within the stationary research-share class}\label{tab:domestic_sensitivity}
\begin{tabular}{@{}lrrrrrr@{}}
\toprule
Scenario & $\kappa$ & $x_R$ (\%) & $P_R$ (\%) & $P_J$ (\%) & $P_E$ (\%) & Contest share (\%)\\
\midrule
Baseline & 0.0750 & 50.00 & 75.00 & 5.83 & 11.79 & 91.39\\
Race target 25\% & 0.0608 & 25.00 & 65.91 & 5.83 & 11.79 & 90.08\\
Race target 75\% & 0.0818 & 75.00 & 79.29 & 5.83 & 11.79 & 91.89\\
State weight $\omega=0.25$ & 0.0860 & 50.00 & 75.00 & 4.56 & 10.58 & 91.46\\
State weight $\omega=0.75^{\dagger}$ & 0.0640 & 50.00 & 75.00 & 7.04 & 12.92 & 91.35\\
Prize share $\pi=0.25$ & 0.0750 & 8.42 & 48.70 & 5.83 & 11.79 & 86.10\\
Prize share $\pi=0.75$ & 0.0750 & 95.00 & 81.42 & 5.83 & 11.79 & 92.12\\
\bottomrule
\end{tabular}
\par\smallskip
\begin{minipage}{\textwidth}\footnotesize
\emph{Notes:} $P_J$ is the joint-profit benchmark among common shares; $P_E$ removes the contest. Contest share is $100(P_R-P_E)/(P_R-P_J)$, a sequential probability-gap decomposition. Race-target cases refit $\kappa$ while holding the physical technology fixed. State-weight cases refit safety yield to the original stopped-clock pause target and refit $\kappa$ to a 50\% race share. Prize cases hold baseline $\kappa$ fixed. The $\pi=0.75$ race reaches the 95\% feasible upper corner. All selected race and no-contest shares pass numerical global-grid/refinement best-response checks; this is not a general uniqueness proof. $^{\dagger}$At $\omega=0.75$, the maintained rising-$V$ condition fails on part of the full share grid, although the reported target and best-response checks remain valid; general monotone comparative statics therefore do not apply to the entire range.
\end{minipage}
\end{table}

\begin{table}[htbp]\centering\small
\caption{Racing at Different Safety Yields}\label{tab:sigma}
\setlength{\tabcolsep}{3pt}
\begin{tabular}{@{}ccccccccc@{}}\toprule
Yield & Pause & Full-speed & \multicolumn{2}{c}{Racing labs} & No contest & \multicolumn{2}{c}{Coordination} & Contest share\\ \cmidrule(lr){4-5}\cmidrule(lr){7-8}
$\sigma$ & risk (\%) & risk (\%) & Research (\%) & Risk (\%) & risk (\%) & Research (\%) & Risk (\%) & of the gap (\%)\\ \midrule
0.065 & 60.0 & 85.7 & 95.0$^{\dagger}$ & 89.1 & 14.8 & 0.01 & 7.4 & 91\\
0.086 & 40.0 & 75.0 & 50.0 & 75.0 & 11.8 & 0.06 & 5.8 & 91\\
0.108 & 25.1 & 63.0 & 61.3 & 66.1 & 9.9 & 0.22 & 4.9 & 92\\
0.129 & 15.3 & 51.1 & 78.1 & 59.1 & 8.7 & 0.55 & 4.3 & 92\\
0.172 & 5.8 & 31.6 & 95.0$^{\dagger}$ & 44.0 & 7.1 & 1.96 & 3.5 & 91\\
\bottomrule\end{tabular}
\par\smallskip\begin{minipage}{\textwidth}\footnotesize
\emph{Notes:} The rows scale the baseline safety yield, 0.086 per month, by 0.75, 1, 1.25, 1.5 and 2. The hazard primitives $h$ and $k$, the contest sensitivity $\kappa$, the prize share $\pi$ and $r$ are those of the baseline and are not refitted. Pause risk is the catastrophe probability after the calibration's 12-month pause at the Automated Coder; full-speed risk is at 50\% research. $^{\dagger}$The race sits at the upper constraint. At the lowest yield $V$ peaks below the upper constraint, so it does not rise over the whole interval; the reported shares pass the numerical best-response checks. The last column is $(P_R-P_E)/(P_R-P_J)$.
\end{minipage}\end{table}

The fitted contest sensitivity makes the imposed 50\% share satisfy \eqref{eq:race_foc}. Domestic hazards and their share derivatives use capability quadrature, checked independently against time-domain integration and finite differences. Numerical best-response searches check the reported equilibria but do not prove general uniqueness.

\subsection{A Longer Takeoff Period}

The labs results keep their direction when the full-speed takeoff period lasts 36 months instead of 18 (Table~\ref{tab:takeoff36}). Thirty-six months is a slow case. In the AI Futures Model's published forecasts for this stretch, the medians run from 9 to 18 months and the 90th percentiles from 3.6 to 15.5 years \citep{lifland2026}. The check keeps the hazard targets, compute growth, compute shares, three labs and the prize share. It refits the safety yield, the research-speed scale and $\kappa$, so racing labs again choose 50 percent research.

At that share the value of a month rises from \tkVRacePctEighteen{} to \tkVRacePctThirtySix{} percent a month, so the fitted $\kappa$ rises. Coordination again cuts research to near zero, \tkXCoopThirtySix{} percent, and catastrophe probability to \tkRiskCoopThirtySix{} percent; the coordinated takeoff period takes \tkTCoopThirtySix{} months rather than \tkTCoopEighteen. Removing the contest still closes most of the gap, \tkGapContestThirtySix{} of \tkGapCoordThirtySix{} points. Under the 12-month deadline, catastrophe probability is \tkLeadCoopRiskThirtySix{} rather than \tkLeadCoopRiskEighteen{} percent.

\begin{table}[htbp]
\centering\small
\caption{Baseline and a 36-Month Takeoff Period}\label{tab:takeoff36}
\begin{tabular}{@{}lrr@{}}
\toprule
& \multicolumn{2}{c}{Full-speed takeoff period $T_0$}\\
\cmidrule(l){2-3}
& 18 months (baseline) & 36 months\\
\midrule
\multicolumn{3}{@{}l}{\emph{Calibration}}\\
\quad Full-speed takeoff period (months) & \tkTakeoffEighteen & \tkTakeoffThirtySix\\
\quad Compute growth over the takeoff period (fold) & \tkComputeFoldEighteen & \tkComputeFoldThirtySix\\
\quad Fitted safety yield $\sigma$ (per month) & \tkSigmaEighteen & \tkSigmaThirtySix\\
\quad Fitted contest sensitivity $\kappa$ (per month) & \tkKappaEighteen & \tkKappaThirtySix\\
\addlinespace
\multicolumn{3}{@{}l}{\emph{Racing, 50\% research}}\\
\quad Catastrophe probability (\%) & \tkRiskRaceEighteen & \tkRiskRaceThirtySix\\
\quad Value of a month (\% a month) & \tkVRacePctEighteen & \tkVRacePctThirtySix\\
\addlinespace
\multicolumn{3}{@{}l}{\emph{Coordination, common research share}}\\
\quad Research share (\%) & \tkXCoopEighteen & \tkXCoopThirtySix\\
\quad Takeoff period (months) & \tkTCoopEighteen & \tkTCoopThirtySix\\
\quad Catastrophe probability (\%) & \tkRiskCoopEighteen & \tkRiskCoopThirtySix\\
\addlinespace
\multicolumn{3}{@{}l}{\emph{No contest, $\kappa=0$}}\\
\quad Research share (\%) & \tkXNoContestEighteen & \tkXNoContestThirtySix\\
\quad Catastrophe probability (\%) & \tkRiskNoContestEighteen & \tkRiskNoContestThirtySix\\
\addlinespace
\multicolumn{3}{@{}l}{\emph{Contest's part of the racing--coordination gap}}\\
\quad Points closed by removing the contest & \tkGapContestEighteen & \tkGapContestThirtySix\\
\quad Average share over both removal orders (\%) & \tkContestShapleyEighteen & \tkContestShapleyThirtySix\\
\addlinespace
\multicolumn{3}{@{}l}{\emph{Best common share, 12-month deadline}}\\
\quad Research share (\%) & \tkLeadCoopXEighteen & \tkLeadCoopXThirtySix\\
\quad Takeoff period (months) & \tkLeadCoopTEighteen & \tkLeadCoopTThirtySix\\
\quad Catastrophe probability (\%) & \tkLeadCoopRiskEighteen & \tkLeadCoopRiskThirtySix\\
\bottomrule
\end{tabular}
\par\smallskip
\begin{minipage}{\textwidth}\footnotesize
\emph{Notes:} Each column refits the model with the full-speed takeoff period $T_0$ set to 18 or 36 months. Held fixed: the 75\% catastrophe probability at full speed, the 40\% probability after a 12-month stopped-clock pause at the start, compute doubling every six months, research ten times as fast at TED-AI, the 50/5/45 compute shares, three labs, the prize share $\pi=0.5$, the discount rate, and the 12-month deadline. Refitted: the safety yield $\sigma$, the research-speed scale $b_0$, and $\kappa$, so that racing labs again choose 50\% research. The gap rows remove the contest from the race: the first counts the points of catastrophe probability this closes, the second averages the contest's share of the racing--coordination gap over the two removal orders.
\end{minipage}
\end{table}
\FloatBarrier

\subsection{Short Takeoff Periods}

Shorter takeoff periods break the 50 percent calibration. With a takeoff period shorter than about \lswSwitchTzero{} months, no $\kappa$ makes 50 percent research a racing equilibrium (Table~\ref{tab:takeoff_range}). A short takeoff period packs the same 75 percent probability into fewer months, so each month at a capability level is more dangerous and speeding up saves more exposure. The value of a month is the hazard added by arriving with less safety knowledge minus the hazard removed by spending fewer months exposed. At 50 percent research the exposure part is \lswExpoNine{} percent a month with a 9-month takeoff period and \lswExpoEighteen{} percent with an 18-month takeoff period. The knowledge part barely changes, \lswKnowNine{} against \lswKnowEighteen{} percent.

With a 9-month takeoff period the value of a month peaks at \lswPeakNine{} percent research and falls beyond it. A $\kappa$ fitted to the 50 percent share then puts the racing hurdle where $V$ is falling, which is a payoff minimum. A lab whose rivals choose 50 percent does better by slowing almost to zero or by racing to the upper constraint. The 50 percent calibration also fails when the pause target exceeds about \lswSwitchPreg{} percent or the state-hazard weight exceeds about \lswSwitchOmega.

Holding $\kappa$ at its baseline value instead keeps a race at every takeoff length (bottom panel of Table~\ref{tab:takeoff_range}). The contest sensitivity describes how the market rewards a head start, so it need not change with the length of the takeoff period. With $\kappa$ held at \lswKappa{} per month, labs race at the 95 percent upper constraint with 9- and 12-month takeoff periods and at \lswXRaceThirtySix{} percent research with a 36-month takeoff period. Shorter takeoff periods make labs race harder because the value of a month is lower. Removing the contest closes \lswHeldFirstMin{} to \lswHeldFirstMax{} percent of the racing--coordination gap at every length, and \lswHeldShapMin{} to \lswHeldShapMax{} percent averaged over both removal orders.

\begin{table}[htbp]
\centering\small
\caption{Takeoff Length with the Contest Sensitivity Refitted or Held}\label{tab:takeoff_range}
\begin{tabular}{@{}lrrrrr@{}}
\toprule
& \multicolumn{5}{c}{Full-speed takeoff period $T_0$ (months)}\\
\cmidrule(l){2-6}
& 9 & 12 & 18 (baseline) & 24 & 36\\
\midrule
\multicolumn{6}{@{}l}{\emph{Not depending on $\kappa$}}\\
\quad Fitted safety yield $\sigma$ (per month) & \lswSigmaNine & \lswSigmaTwelve & \lswSigmaEighteen & \lswSigmaTwentyFour & \lswSigmaThirtySix\\
\quad Coordinated research share (\%) & \lswXCoopNine & \lswXCoopTwelve & \lswXCoopEighteen & \lswXCoopTwentyFour & \lswXCoopThirtySix\\
\quad Coordinated takeoff period (months) & \lswTCoopNine & \lswTCoopTwelve & \lswTCoopEighteen & \lswTCoopTwentyFour & \lswTCoopThirtySix\\
\quad Catastrophe probability, coordination (\%) & \lswRiskCoopNine & \lswRiskCoopTwelve & \lswRiskCoopEighteen & \lswRiskCoopTwentyFour & \lswRiskCoopThirtySix\\
\quad Catastrophe probability, no contest (\%) & \lswRiskNoContestNine & \lswRiskNoContestTwelve & \lswRiskNoContestEighteen & \lswRiskNoContestTwentyFour & \lswRiskNoContestThirtySix\\
\addlinespace
\multicolumn{6}{@{}l}{\emph{$\kappa$ refitted so that racing labs choose 50\% research}}\\
\quad Fitted $\kappa$ (per month) & \lswFitKappaNine & \lswFitKappaTwelve & \lswFitKappaEighteen & \lswFitKappaTwentyFour & \lswFitKappaThirtySix\\
\quad Catastrophe probability, racing (\%) & \lswFitRiskNine & \lswFitRiskTwelve & \lswFitRiskEighteen & \lswFitRiskTwentyFour & \lswFitRiskThirtySix\\
\quad Gap closed by removing the contest (\%) & \lswFitFirstNine & \lswFitFirstTwelve & \lswFitFirstEighteen & \lswFitFirstTwentyFour & \lswFitFirstThirtySix\\
\quad Average over both removal orders (\%) & \lswFitShapNine & \lswFitShapTwelve & \lswFitShapEighteen & \lswFitShapTwentyFour & \lswFitShapThirtySix\\
\addlinespace
\multicolumn{6}{@{}l}{\emph{$\kappa$ held at its baseline value, \lswKappa{} per month}}\\
\quad Racing research share (\%) & \lswHeldXNine & \lswHeldXTwelve & \lswHeldXEighteen & \lswHeldXTwentyFour & \lswHeldXThirtySix\\
\quad Racing takeoff period (months) & \lswHeldTNine & \lswHeldTTwelve & \lswHeldTEighteen & \lswHeldTTwentyFour & \lswHeldTThirtySix\\
\quad Catastrophe probability, racing (\%) & \lswHeldRiskNine & \lswHeldRiskTwelve & \lswHeldRiskEighteen & \lswHeldRiskTwentyFour & \lswHeldRiskThirtySix\\
\quad Gap closed by removing the contest (\%) & \lswHeldFirstNine & \lswHeldFirstTwelve & \lswHeldFirstEighteen & \lswHeldFirstTwentyFour & \lswHeldFirstThirtySix\\
\quad Average over both removal orders (\%) & \lswHeldShapNine & \lswHeldShapTwelve & \lswHeldShapEighteen & \lswHeldShapTwentyFour & \lswHeldShapThirtySix\\
\bottomrule
\end{tabular}
\par\smallskip
\begin{minipage}{\textwidth}\footnotesize
\emph{Notes:} Each column refits the safety yield $\sigma$ and the research-speed scale $b_0$ to its full-speed takeoff period, holding the 75\% and 40\% probability targets, compute growth, the 50/5/45 compute shares, three labs, the prize share $\pi=0.5$ and the discount rate. Coordination and the no-contest benchmark do not depend on $\kappa$. In the middle panel $\kappa$ is refitted so that 50\% research is the racing equilibrium; with 9- and 12-month takeoff periods no $\kappa$ does this (dashes). In the bottom panel $\kappa$ keeps its 18-month value and the racing share is the symmetric equilibrium; 95\% is the upper constraint. The gap rows give the share of the racing--coordination gap in catastrophe probability that removing the contest closes, first with the contest removed first and then averaged over both removal orders.
\end{minipage}
\end{table}

\subsection{Other Parameter Checks}

\begin{figure}[htbp]
\centering
\includegraphics[width=\textwidth]{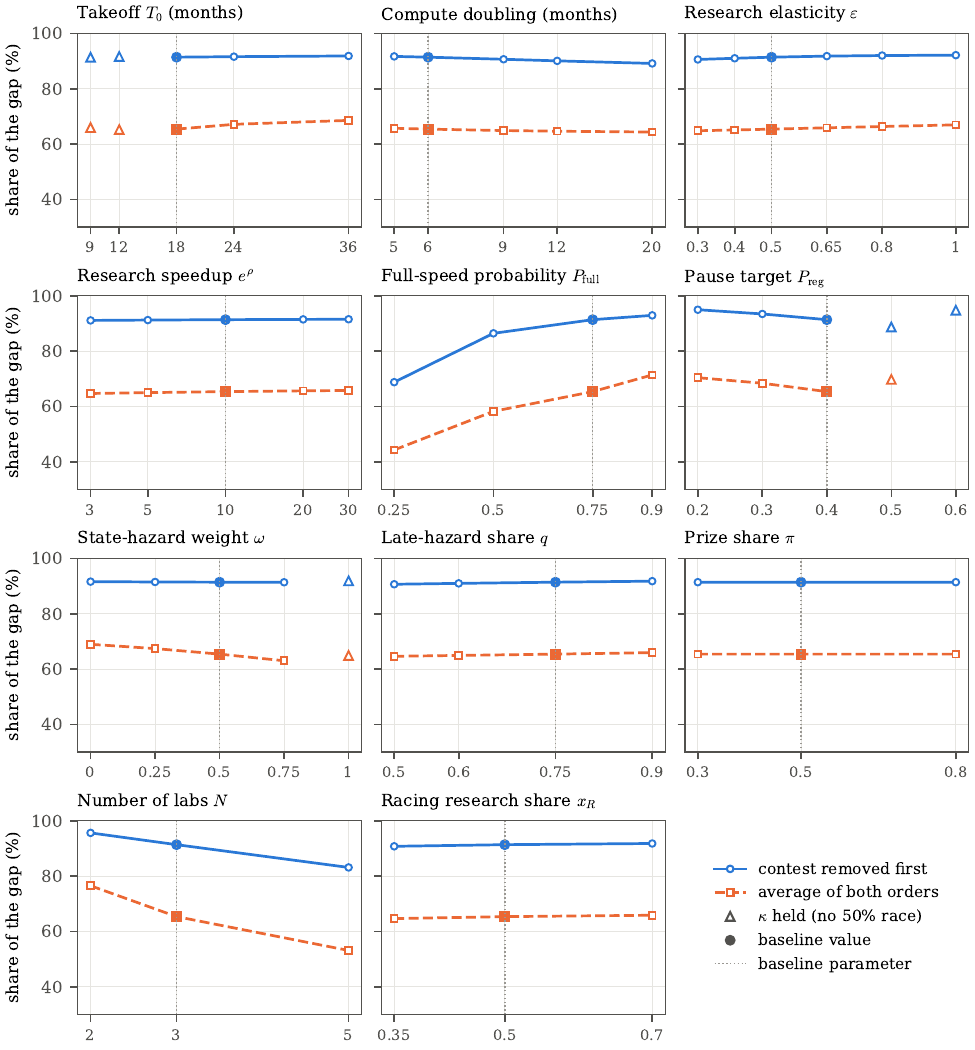}
\caption{The contest's share of the racing--coordination gap, one parameter at a time. Each panel varies one parameter and refits the safety yield, the hazard scale and $\kappa$, so that racing labs choose 50\% research (the last panel varies that share). Circles: the contest removed first; squares: the average over both removal orders. Triangles mark values where no $\kappa$ makes 50\% research an equilibrium; there $\kappa$ keeps its baseline value and labs race at the 95\% upper constraint. At a 60\% pause target the average is not shown, because without the hazard externality the game has no symmetric equilibrium. Filled markers and dotted lines show the baseline.}\label{fig:labs_sweep}
\end{figure}

Figure~\ref{fig:labs_sweep} varies each parameter one at a time, refitting the safety yield, the hazard scale and $\kappa$ each time. Removing the contest closes between \lswSweepMin{} and \lswSweepMax{} percent of the gap across these cases.

The share falls below ninety percent in five cases. Two of them lower the full-speed probability, to 50 percent (\lswPfullHalf{} percent of the gap) and to 25 percent (\lswPfullQuarter{} percent). The others are five labs (\lswNFive{} percent), compute doubling every 20 months (\lswDoublingTwenty{} percent), and a 50 percent pause target (\lswPregHalf{} percent, $\kappa$ held). The average over both removal orders falls below one half only when the full-speed probability is 25 percent (\lswShapPfullQuarter{} percent). Below a full-speed probability of about \hzFloor{} percent, no contest sensitivity supports the imposed 50 percent racing share. The value of a month at that share is proportional to cumulative hazard, so it falls below the no-contest hurdle $Nr=\rrNoContestHurdle\%$ a month: it is \hzVTen\% at a full-speed probability of 10 percent and \hzVFive\% at 5 percent. Discounting and the hazard externality alone then push labs past the imposed share. Above the limit, the contest's share of the gap with the contest removed first falls from \lswPfullQuarter{} percent at a full-speed probability of 25 percent to \hzFirstTwenty{} at 20 and \hzFirstFifteen{} at 15, and reaches zero at the limit.

Varying the safety share or the learning multiplier leaves every number unchanged. The labs game holds safety at its reference share, and the safety yield is refitted at that share. Varying the reference research share $x_0$ with labs racing at $x_0$ also leaves every number unchanged, so the last panel varies the racing share instead.

Parameter provenance, dependency hashes, first-order-condition residuals, and integration comparisons for these results are available from the author.

\FloatBarrier
\section{Rival Responses near Symmetry}\label{app:cross}

Whether rivals slow down when one lab restrains depends on a cross partial. The survival and discounting terms of $\ln U_i$ separate across labs, so only the prize term $\ln(B+\Pi p_i)$ has a cross partial. Its derivative in $x_i$ is $\kappa\tau_iq(p_i)$, where $q(p)=\Pi p(1-p)/(B+\Pi p)$, and $\partial p_i/\partial x_j=-\kappa\tau_jp_ip_j$. So for different labs $i,j$,
\begin{equation}
\frac{\partial^2\ln U_i}{\partial x_i\partial x_j}
=-\kappa^2\tau_i\tau_jp_ip_jq'(p_i),\qquad
q'(p)=\frac{\Pi[B(1-2p)-\Pi p^2]}{(B+\Pi p)^2}.
\label{eq:contest-cross-partial}
\end{equation}
The cross partial says whether a rival's faster research raises or lowers a lab's marginal gain from speed. The term in front of $q'$, $-\kappa^2\tau_i\tau_jp_ip_j$, is negative, so the cross partial has the opposite sign to $q'$. At symmetry $p_i=1/N$, it is positive when $\Pi/B>N(N-2)$, negative when smaller, and zero at equality. Under a strict local second-order condition these signs give the local best-response direction. At the baseline, $\Pi=3B$ and $N=3$, so $q'(1/3)$ is proportional to $B/3-\Pi/9=0$.

\FloatBarrier
\section{Proofs of the Propositions}\label{app:proofs}

\begin{proof}[Proof of Proposition~\ref{prop:race}]
The log payoff derivative is
\[
\frac{\partial\ln U_n}{\partial x_n}
=-\frac{H'(x_n)}N+r\tau(x_n)
+\frac{\Pi}{\Pi p_n+B}\frac{\partial p_n}{\partial x_n},
\qquad
\frac{\partial p_n}{\partial x_n}
=\kappa\tau(x_n)p_n(1-p_n).
\]

\begin{itemize}
\item At a symmetric Nash equilibrium every lab chooses $x_R$, so each wins with probability $p_n=1/N$ and the last term is $\kappa\tau(x_R)(N-1)\pi/N$. Setting the derivative to zero and multiplying by $N/\tau$ gives equation~\eqref{eq:race_foc}.
\item At a common share the joint objective is $e^{-H(x)}e^{-rT(x)}(\Pi+NB)$; its log derivative is $-H'(x)+r\tau(x)$, giving equation~\eqref{eq:coord_foc}. 
\item Setting $\kappa=0$ makes $\partial p_n/\partial x_n=0$, so the last term vanishes and the first-order condition gives equation~\eqref{eq:externality_foc}. The ordering follows from increasing $V$. At the upper boundary, a necessary racing condition is $V(1-s_0)\le Nr+\kappa(N-1)\pi$.
\end{itemize}
\end{proof}

\begin{proof}[Proof of Proposition~\ref{prop:labs}]
Treating $N$ as continuous,
\[
\frac{\partial}{\partial N}\frac{(N-1)\Pi}{\Pi+NB}
=\frac{\Pi(\Pi+B)}{(\Pi+NB)^2}>0.
\]
The derivative of $Nr$ is $r>0$. Interior shares rise with the hurdle by equation~\eqref{eq:race_foc}; at the upper constraint they can remain at the same share, as in the four- and five-lab cases.
\end{proof}

\begin{proof}[Proof of Proposition~\ref{prop:publish}]
A lab's payoff is $U_n^P=e^{-H(z)}e^{-rT(z)}(B+\Pi/N)$. The contest probability is constant, and
\[
\frac{\partial\ln U_n^P}{\partial x_n}
=\frac{\partial z}{\partial x_n}[-H'(z)+r\tau(z)]
=\frac1N[-H'(z)+r\tau(z)].
\]
The interior condition is $V(z)=r$. Joint profits have the same path-dependent factor, multiplied by $\Pi+NB$, so the aggregate condition matches common-path coordination.
\end{proof}

\FloatBarrier
\newpage
\section{Notation}\label{app:notation}
\begin{table}[htbp]
\centering\small
\caption{Notation, in order of appearance}\label{tab:notation}
\begin{tabular}{@{}p{.27\textwidth}p{.68\textwidth}@{}}
\toprule
Symbol & Meaning\\
\midrule
$K(t),g_K,\gamma$ & Installed compute; its growth rate; growth of research capacity, $\varepsilon g_K$\\
$i,x,s$ & Shares for customer inference, capability research, and safety research\\
$F,\phi(F),\varepsilon$ & Capability (0 Automated Coder, 1 TED-AI); research productivity of a model of capability $F$; research elasticity\\
$S,\sigma(F),y(s)$ & Safety knowledge; safety yield; learning multiplier\\
$\theta,\rho,\psi,\alpha$ & Calibration forms $\phi(F)=\theta e^{\rho F}$ and $y(s)=\psi+(1-\psi)(s/s_0)^\alpha$\\
$h(F),k(F),\lambda(t),H,P$ & State and step hazard; hazard rate; cumulative hazard; catastrophe probability\\
$T(x),\tau(x),V(x)$ & Arrival date at research share $x$; months saved per unit share; value of a month, the hazard per month saved, $H'(x)/\tau(x)$\\
$t_x(F),f_x(F)$ & Date at which the path at share $x$ reaches capability $F$; its speed $\dot F$ there\\
$N,r,B,\Pi,\pi$ & Number of labs; monthly discount rate; enterprise value; prize; prize share $\Pi/(\Pi+NB)$\\
$p_n,\kappa$ & Probability lab $n$ wins the commercial prize; contest sensitivity: each month of head start multiplies a lab's odds by $e^\kappa$\\
$U_n,U_{\text{joint}},U_n^E$ & Lab $n$'s payoff when labs race; the labs' joint payoff at a common share; lab $n$'s payoff with no contest ($\kappa=0$)\\
$x_R,x_J,x_E$ & Racing share (symmetric equilibrium); coordinated share (joint-profit optimum among common shares); no-contest share (symmetric equilibrium at $\kappa=0$)\\
$U_n^A,x_A,P_A$ & Lab $n$'s payoff, the symmetric equilibrium share, and the catastrophe probability when each lab counts all of the hazard its own research adds but keeps the contest\\
$z$ & Published effective input, the average contribution\\
\bottomrule
\end{tabular}
\end{table}

\end{document}